\documentclass[pdflatex,sn-mathphys-num, iicol]{sn-jnl}

\usepackage{amsmath}
\usepackage{amssymb}
\usepackage{booktabs}
\usepackage{csquotes}
\usepackage[T1]{fontenc}
\usepackage{siunitx}
\usepackage{scalefnt}
\usepackage[normalem]{ulem}
\usepackage{listings}

\usepackage{color}
\usepackage[dvipsnames]{xcolor}
\usepackage{multirow}
\usepackage{placeins}
\usepackage{cleveref}
\newcommand*\mathd{\mathop{}\!\mathrm{d}}

\newcommand{\Matrix}{\textsc{Matrix}}
\newcommand{\Munich}{\textsc{Munich}}
\newcommand{\PineAPPL}{\textsc{PineAPPL}}
\newcommand{\fastNLO}{\textsc{fastNLO}}
\newcommand{\APPLgrid}{\textsc{APPLgrid}}
\newcommand{\MatrixHawaii}{\textsc{Matrix Hawaii}}
\newcommand{\ATLASCitation}{\cite{ATLAS:2016nqi}}
\newcommand{\LHCbCitation}{\cite{LHCb:2015okr}}
\newcommand{\CMSCitation}{\cite{CMS:2015rld}}
\newcommand{\ATLAS}{ATLAS~\ATLASCitation{}}
\newcommand{\LHCb}{LHCb~\LHCbCitation{}}
\newcommand{\CMS}{CMS~\CMSCitation{}}
\newcommand{\rcut}{\ensuremath{r_\mathrm{cut}}}
\newcommand{\rcutmin}{\ensuremath{r_\mathrm{cut,min}}}
\newcommand{\rcutmax}{\ensuremath{r_\mathrm{cut,max}}}
\newcommand{\rcuti}{r_\mathrm{cut}^{(i)}}
\newcommand{\rcutj}{r_\mathrm{cut}^{(j)}}

\begin{document}  
\title[MATRIX HAWAII: PineAPPL interpolation grids with MATRIX]{MATRIX HAWAII: PineAPPL interpolation grids with MATRIX}

\author[1]{\fnm{Simone} \sur{Devoto}}\email{simone.devoto@ugent.be}
\author[2]{\fnm{Tomas} \sur{Je\v{z}o}}\email{tomas.jezo@uni-muenster.de}
\author[3]{\fnm{Stefan} \sur{Kallweit}}\email{stefan.kallweit@physik.uzh.ch}
\author[4]{\fnm{Christopher} \sur{Schwan}}\email{christopher.schwan@uni-wuerzburg.de}

\affil[1]{\orgdiv{Department of Physics and Astronomy}, \orgname{Ghent University}, \orgaddress{\city{Ghent}, \postcode{9000}, \country{Belgium}}}
\affil[2]{\orgdiv{Institut  für  Theoretische  Physik}, \orgname{Universität Münster}, \orgaddress{\city{Münster}, \postcode{48149}, \country{Germany}}}
\affil[3]{\orgdiv{Physik Institut}, \orgname{Universit\"at Z\"urich}, \orgaddress{\city{ Z\"urich}, \postcode{8057}, \country{Switzerland}}}
\affil[4]{\orgdiv{Institut für Theoretische Physik und Astrophysik}, \orgname{Universit\"at W\"urzburg}, \orgaddress{\city{Würzburg}, \postcode{97074}, \country{Germany}}}

\flushbottom

\abstract{
We present an interface between \PineAPPL{} and \Matrix{}, which allows fully differential cross sections to be calculated in the form of interpolation grids, accurate at next-to-next-to-leading order (NNLO) in QCD and next-to-leading order in electroweak~(EW) theory.
This interface is the first publicly available tool to calculate interpolation grids at NNLO QCD accuracy for a wide set of processes.
Interpolation grids provide the functionality to compute predictions for arbitrary parton distribution functions (PDFs) as well as PDF uncertainties without the need to repeat the actual calculation.
Another important application of the these grids is to perform global analyses of PDFs using exact NNLO calculations instead of $K$-factors, which have several drawbacks.
This exact treatment of NNLO corrections is also an important prerequisite for fitting PDFs at next-to-next-to-next-to-leading order level with reliable uncertainties.
The new version of the \Matrix{} code interfaced to \PineAPPL{}, as well as the grids produced for this publication, are available on the \Matrix{} website and on {\tt PloughShare}, respectively.
}

\maketitle

\section{Introduction}

As we near the high-luminosity phase of the Large Hadron Collider (LHC), we approach an unprecedented level of precision and accuracy for experimental measurements of all standard model (SM) processes.
To learn the most from data--theory comparisons, matching this level of precision in the corresponding predictions is required.
Especially for well-known processes like Drell--Yan production the experimental uncertainties already rival those from theory predictions.

In the absence of any significant tension between theory and experiments, the measurements and corresponding predictions are used to improve knowledge of proton structure encoded in parton distribution functions (PDFs).
These are a key ingredient for all other predictions for hadronic collisions, and as we strive for unprecedented precision, the refinement of PDFs becomes paramount.
Achieving higher precision makes PDF determinations more computationally costly, but so-called {\em interpolation grids}~\cite{Kluge:2006xs,Carli:2010rw,Ball:2010de,Carrazza:2020gss} mitigate this computational challenge to a large extent.
In an interpolation grid predictions are stored in a PDF-independent way, so that the most time-consuming part of the computation is only performed once and convolutions with arbitrary PDFs can be performed very quickly a-posteriori.
The use of interpolation grids in global analyses of PDFs is by now widespread at next-to-leading order (NLO) QCD and is getting adapted~\cite{Ball:2011mu,ATLAS:2018owm,H1:2021xxi} as soon as they are available at next-to-next-to-leading order (NNLO) QCD~\cite{Britzger:2019kkb,Britzger:2022lbf,Czakon:2017dip, Czakon:2019yrx, Cruz-Martinez:2025ffa}.
However, the availability of grids at NNLO QCD is currently limited.
Moreover, there is currently no publicly available program with a documented public interface for the generation of
NNLO QCD interpolation grids for LHC processes.
The goal of this publication is to address this issue by providing a tool capable of producing grids based on a combination of \Matrix{}~\cite{Grazzini:2017mhc} and \PineAPPL{}~\cite{Carrazza:2020gss,christopher_schwan_2024_10698034}, named \MatrixHawaii\footnote{\MatrixHawaii{} is the abbreviation of ``\Munich -- the MUlti-chaNnel Integrator at swiss (CH) precision -- Automates qT-subtraction and Resummation to Integrate X-sections, Handling Automation With Additional pIneappl -- Pineappl Is Not an Extension of APPLgrid -- Interpolation-grids''}.
\Matrix{} provides predictions for a set of processes that are indispensable for realistic PDF determinations including Drell--Yan off-shell $Z$/$W$-boson production and $t\bar{t}$ production up to NNLO QCD+NLO EW~\cite{Grazzini:2019jkl} and NNLO QCD~\cite{Catani:2019iny,Catani:2019hip}, respectively.
Complementary to that, \PineAPPL{} provides the means to grid those predictions, notably for the first time also implementing a way to combine QCD and EW corrections.
Providing this tool\footnote{This idea was already explored in Ref.~\cite{Garzelli:2023rvx} in a private extension of \Matrix{}, limited to $t\bar{t}$ production and without a generalized extrapolation procedure (see \Cref{sec:rcut-parameter-dependence}).} is a crucial step towards making precise predictions in the form of interpolation grids available to a broader community.

In most modern PDF extractions at NNLO, the grids are often missing and therefore a $K$-factor approximation is applied instead, in which the NNLO QCD predictions are obtained by multiplying the NLO QCD predictions with $K$-factors, defined as the ratio of NNLO to NLO.
Whereas these $K$-factors are usually computed on a bin-by-bin basis, they are assumed to be independent of both the PDFs with which they were generated and the partonic channels.
The reliability of this approximation can only be assessed by comparing a fit that uses $K$-factors to the same fit that instead uses full NNLO calculations.
In the absence of such fit, we design a proxy to gain some insight on the impact of using full NNLO calculations instead of relying on $K$-factors both on the central PDFs and the uncertainty bands.

This paper is organized as follows: in \Cref{sec:interface} we briefly describe \Matrix{} and \PineAPPL{}, and detail the new interface.
In \Cref{sec:validation} we validate our implementation by producing grids for a small selection of LHC measurements, and comparing them to the results provided by \Matrix{} directly.
In \Cref{sec:applications} we showcase possible applications of our interface, including an investigation of the difference between using full NNLO QCD predictions and $K$-factors in PDF fits.
In \Cref{sec:conclusions} we present our conclusions.
Finally, in \Cref{app:manual} we describe the installation of this toolchain and how to use it, and in \Cref{app:extrapolation} we describe technical aspects of the implementation that allow us to control the power corrections at the level of interpolation grids.

\section{\PineAPPL{} interface to \Matrix{}\label{sec:interface}}

An interpolation grid is a representation of a (multi-)differential distribution independent of the PDFs and the strong coupling, corresponding  to a binned histogram produced by a Monte Carlo (MC) integrator. 
The advantage of such a representation is that the time-consuming integration of the matrix elements needs to be performed only once; after their generation the grids can be convolved with any PDF set.
This convolution with PDFs is an operation that is very fast, usually taking only a fraction of a second.
Building on this basic operation, interpolation grids offer further fast operations and applications:
\begin{itemize}
\item \emph{PDF uncertainties}.
Interpolation grids naturally support the calculation of PDF uncertainties.
Without grids, the calculation of PDF uncertainties on-the-fly in NNLO predictions may be available~\cite{Campbell:2019dru,Monni:2019whf}, but still poses a significant computational challenge due to memory limitations and the speed of IO operations.
\item \emph{Channel sizes}.
Since interpolation grids store the results for each partonic channel separately, one can study the size of each channel and how it changes in different bins for different PDF sets.
This allows us to better understand the process itself and the differences due to the choice of a PDF set.
\item \emph{PDF fits}.
Since interpolation grids are independent of PDFs, they are an important ingredient of PDF determinations.
As soon as a prediction for a measurement is available as an interpolation grid, it can be relatively easily included in a PDF fit.
\end{itemize}
Yet another advantage of interpolation grids is that, once they are generated, the corresponding MC integrators are no longer needed to perform these operations.
Therefore, they can be understood as an efficient exchange format, between prediction providers and (possibly different) users, making grids the ideal representation to store predictions.
Moreover, when PDFs become more accurate and precise, interpolation grids can be used to easily redo computations with the updated PDFs.

\subsection{Representation of an interpolation grid}

For proton--proton collisions, the factorisation theorem allows us to write cross sections (up to terms of higher twist) in the following way,
\begin{align}
\sigma = &\sum_{ab} \sum_{nm} \int_0^1 \mathrm{d} x_1 \int_0^1 \mathrm{d} x_2 f_a^\mathrm{p} (x_1, Q^2) f_b^\mathrm{p} (x_2, Q^2) 
\\\nonumber&\times
\int \mathrm{d} \mathrm{LIPS} \, \alpha_S^n (Q^2) \alpha^m \sigma_{ab}^{(n,m)} (x_1, x_2, Q^2) \,,
\label{eq:factorization-formula}
\end{align}
where $f_{a/b}^\mathrm{p} (x_1, x_2)$ is the proton PDF for parton $a/b$ with momentum fraction $x_1$, $x_2$, evaluated at scale $Q^2$, $\mathrm{d} \mathrm{LIPS}$ the Lorentz-invariant phase space differential and $\alpha_S (Q^2)$ the strong coupling evaluated with the renormalisation scale set to $Q^2$, and $\sigma_{ab}^{(n,m)}$ the partonic cross section for the partons $a$ and $b$ at the $n$-th and $m$-th order in the QCD and EW coupling, respectively.
The functional form of the partonic cross sections $\sigma_{ab}$ in terms of $x_1$, $x_2$ and $Q^2$ is unknown in general, which makes it impossible to change the PDFs a-posteriori without recomputing \cref{eq:factorization-formula}.
To circumvent this problem, we evaluate the PDFs in their variables $x$ and $Q^2$ on a two-dimensional grid of node points $\{ x^{(j)} \}_{j=1}^N$ and $\{ Q^2_k \}_{k=1}^M$.
This grid is used to interpolate the PDFs between the node points, which are chosen in such a way that the interpolation error is kept small.
Then \cref{eq:factorization-formula} can be approximated by
\begin{align}
\sigma \approx &\sum_{ab} \sum_{nm} \sum_{ijk} f_a^\mathrm{p} (x_1^{(i)}, Q^2_k) f_b^\mathrm{p} (x_2^{(j)}, Q^2_k) 
\\\nonumber&\times
\alpha_S^n (Q^2_k) \alpha^m \Sigma_{ab}^{(n,m)} (x_1^{(i)}, x_2^{(j)}, Q^2_k) \text{,}
\end{align}
where the set of numbers
\begin{equation}
\left\{ \alpha^m \Sigma_{ab}^{(n,m)} (x_1^{(i)}, x_2^{(j)}, Q^2_k) \right\}_{ab,ijk,nm}
\end{equation}
represent the interpolation grid.
Note that powers of $\alpha$, which is usually chosen constant w.r.t.\ $Q^2$, are part of the interpolation grid, and thus the index $m$ is there only to distinguish perturbative orders with the same power of $\alpha_S$.
Effectively, for scattering of partons $a$ and $b$, an interpolation grid is a set of partonic cross sections, differential in $x_1$, $x_2$ and $Q^2$, and split according to the orders of QCD and EW couplings, $n$ and $m$.

\PineAPPL{}~\cite{Carrazza:2020gss,christopher_schwan_2024_10698034} is an implementation of the idea outlined above.
It was written to support arbitrary perturbative contributions/corrections in powers of the strong and electroweak couplings, which distinguishes it from \fastNLO{}~\cite{Kluge:2006xs,Wobisch:2011ij,Britzger:2012bs} and \APPLgrid{}~\cite{Carli:2010rw}.
Furthermore, it offers interfaces in various programming languages, namely C, C++, Fortran, Python and Rust, and comes with the command-line program \texttt{pineappl} to convolve the grids and to perform further checks and analyses.

\subsection{\MatrixHawaii{}}

\Matrix{} is a public computational framework which allows single- and double-differential distributions to be evaluated at NNLO QCD accuracy for a wide, and constantly increasing, set of processes.
The latest additions to the class of supported processes are top-quark pair production~\cite{Catani:2019iny,Catani:2019hip} and triphoton production~\cite{Kallweit:2020gcp} in version 2.1 of the code ({\tt v2.1}).
Since the release of version 2 ({\tt v2}) also NLO EW corrections~\cite{Grazzini:2019jkl} have been added to the framework for most single-boson and diboson processes, as well as NLO QCD corrections to loop-induced gluon channels in massive diboson production~\cite{Grazzini:2018owa,Grazzini:2020stb}.

Recent developments further extended the scope of \Matrix, and, while they are currently not available in the public release of the framework yet, they are to be eventually included in future versions of the code.
These include mixed QCD--EW corrections for the neutral-current~\cite{Bonciani:2021zzf,Armadillo:2024ncf} and charged-current~\cite{Buonocore:2021rxx} Drell--Yan processes.
Similarly, by using different approximations in order to make up for the lack of the exact two-loop amplitudes, several associated heavy-quark production processes have been computed within \Matrix{}: Higgs-boson production in association with a top pair~\cite{Catani:2021cbl,Catani:2022mfv,Devoto:2024nhl} and $W$-boson production associated with either a massive bottom pair~\cite{Buonocore:2022pqq} or a top pair~\cite{Buonocore:2023ljm}. In Refs.~\cite{Buonocore:2023ljm,Devoto:2024nhl} the \MatrixHawaii{} interface presented here has already been applied to predict PDF and $\alpha_S$ uncertainties at NNLO QCD accuracy plus the full tower of sub-leading LO and NLO contributions (sub-leading, in terms of powers of $\alpha_S$, Born production modes as well as all QCD, EW and mixed corrections at NLO). Moreover, in Ref. \cite{Chiefa:2025loi} \MatrixHawaii{} has been already used by the NNPDF collaboration to perform a thorough comparison between experimental data not yet employed in PDF determinations and state-of-the-art theoretical predictions.

The core of \Matrix{} is the MC integrator \Munich, which uses the dipole subtraction method~\cite{Catani:1996jh,Catani:1996vz,Catani:2002hc,Kallweit:2017khh,Dittmaier:1999mb,Dittmaier:2008md,Gehrmann:2010ry,Schonherr:2017qcj} for the evaluation of all NLO-like contributions.
For NNLO computations, \Matrix{} implements the $q_\mathrm{T}$-subtraction formalism~\cite{Catani:2007vq,Bonciani:2015sha,Catani:2019iny,Catani:2019hip,Catani:2023tby} to handle and cancel infrared divergences.
All required amplitudes up to the one-loop level are obtained from \textsc{OpenLoops}~\cite{Cascioli:2011va, Buccioni:2017yxi,Buccioni:2019sur} or \textsc{Recola}~\cite{Actis:2012qn,Actis:2016mpe,Denner:2017wsf}, while the two-loop amplitudes are either provided in the form of numerical grids, analytic expressions, or by linking against external amplitude codes.

By itself, \PineAPPL{} cannot create any predictions, for which it needs a Monte Carlo integrator.
At the same time, \Matrix{} is lacking the functionality to generate interpolation grids.
In this paper we describe an interface between the two codes, named \MatrixHawaii{}, which enables the computation of interpolation grids at NNLO QCD and NLO EW accuracy.

One important development was the generalization of the extrapolation procedure to interpolation grids, which will be described in Appendix \ref{app:extrapolation}.
This is necessary to remove the dependence on the technical slicing parameter used by \Matrix{} in NNLO calculations.

\section{Validation\label{sec:validation}}

In order to establish reliability of our approach, let us first describe the validation of the predictions produced by \MatrixHawaii{} by comparing the exact results delivered directly by \Matrix{} to those from convolving the produced \PineAPPL{} interpolation grids with the same PDF set used by \Matrix{}.

We perform this validation on a set of predictions that, together with their measurements, play an important role in current PDF determinations.
We select the measurements of $\gamma/Z$ and $W$ (pseudo)rapidities by \ATLAS{} and \LHCb{}, both at \SI{7}{\tera\electronvolt}, and the measurement of $t\bar{t}$ transverse momentum, rapidity and mass of top(-pairs) by CMS at \SI{8}{\tera\electronvolt}~\CMSCitation{}.
In \Cref{tab:datasets} we list the cuts that were applied in the measurements.
For the predictions, we choose the values of the renormalisation and factorisation scales as \mbox{$\mu = \mu_\mathrm{R} = \mu_\mathrm{F} = m_i$}, where we set \mbox{$m_i = m_Z$} for Drell--Yan (DY) lepton-pair production and \mbox{$m_i = m_W$} for single-lepton production.
For top--anti-top production, we use a dynamical scale, \mbox{$\mu = \mu_\mathrm{R} = \mu_\mathrm{F} = (H_\mathrm{T,t} + H_\mathrm{T,\bar{t}})/4$}, where
\begin{equation}
H_\mathrm{T,\mathrm{t}} = \sqrt{m_t^2 + p_{\mathrm{T}, t}^2}\,,
\end{equation}
and $H_\mathrm{T,\bar{t}}$ is defined correspondingly.
We use the following choices of masses and widths of all the relevant particles:
\begin{align}
  \nonumber
  m_Z &= \SI{91.1876}{\giga\electronvolt}\,, \qquad & \Gamma_Z &= \SI{2.4952}{\giga\electronvolt}\,, \\\nonumber
  m_W &= \SI{80.385}{\giga\electronvolt}\,,         & \Gamma_W &= \SI{2.0854}{\giga\electronvolt}\,, \\\nonumber
  m_H &= \SI{125}{\giga\electronvolt}\,,            & \Gamma_H &= \SI{4.07468}{\mega\electronvolt}\,, \\
  m_t &= \SI{173.2}{\giga\electronvolt}\,,
\end{align}
and employ the complex-mass scheme~\cite{Denner:2019vbn,Denner:1999gp,Denner:2005fg} throughout.
We use various PDF sets to evaluate our predictions, all provided by the LHAPDF interface~\cite{Buckley:2014ana}.
Similarly, the value of $\alpha_S$, at a given scale $Q$, is taken from the corresponding PDF set with \mbox{$\alpha_S(m_Z) = 0.118$}.

We consider a trivial Cabibbo--Kobayashi--Maskawa~(CKM) matrix in general, apart from DY single-lepton production processes, where the non-diagonal CKM matrix elements are expected to have a non-negligible impact, in particular in the context of PDFs.%
\footnote{Note that all other processes discussed in this paper are independent of the CKM matrix elements when only QCD corrections are considered. For the treatment of the CKM matrix in the context of EW corrections, we refer the reader to \Cref{sec:interpolation-grids-at-nnlo-qcd-nlo-ew}.} 
In this case, we use the following numeric values~\cite{ParticleDataGroup:2024cfk}:
\begin{align}
|V_{\rm CKM}|&=
\left(\begin{array}{ccc}
|V_{ud}| & |V_{us}| & |V_{ub}| \\
|V_{cd}| & |V_{cs}| & |V_{cb}| \\
|V_{td}| & |V_{ts}| & |V_{tb}| \\
\end{array}\right)
\\\nonumber&
=
\left(\begin{array}{ccc}
0.97435 & 0.22501 & 0.003732 \\
0.22487 & 0.97349 & 0.04183 \\
0.00858 & 0.04111 & 0.999118 \\
\end{array}\right)
\end{align}

\begin{table*}
\centering
\begin{tabular}{lcc}
  \toprule
  ATLAS \SI{7}{\tera\electronvolt}~\ATLASCitation{} & $\gamma^*/Z$ (76541.v1, Table 12) & $W^{\pm}$ (76541.v1, Tables 9 and 10) \\
  \midrule
  Observable & $\mathd \sigma / |\mathd y_{l^+l^-}|$ & $\mathrm{d} \sigma / \mathrm{d} |\eta_{\ell^\pm}|$ \\
  \multirow{4}{*}{Cuts} & $p_{T,\ell} > \SI{20}{\giga\electronvolt}$ & $p_{T,{\ell^\pm}} > \SI{25}{\giga\electronvolt}$ \\
                        & $|\eta_{\ell^\pm}| < 2.5$ & $|\eta_{\ell^\pm}| < 2.5$ \\
                        & $\SI{66}{\giga\electronvolt} < m_{\ell^+\ell^-} < \SI{116}{\giga\electronvolt}$ & $p_{\mathrm{T},\text{miss}} > \SI{25}{\giga\electronvolt}$ \\
                        & & $m_{\mathrm{T},W^\pm} \ge \SI{40}{\giga\electronvolt}$  \\
  \bottomrule \\
  \toprule
  LHCb \SI{7}{\tera\electronvolt}~\LHCbCitation{} & $\gamma^*/Z$ (2114.v1, Table 1) & $W^{\pm}$ (2114.v1, Table 4) \\
  \midrule
  Observable & $\mathrm{d} \sigma / \mathrm{d} y_{l^+l^-}$ & $\mathrm{d} \sigma / \mathrm{d} \eta_{l^\pm}$ \\
  \multirow{3}{*}{Cuts} & $p_{\mathrm{T},\ell} > \SI{20}{\giga\electronvolt}$ & $p_{\mathrm{T},{\ell^\pm}} > \SI{20}{\giga\electronvolt}$ \\
                        & $2 < \eta_{\ell^\pm} < 4.5$ & $2 < \eta_{\ell^\pm} < 4.5$ \\
                        & $\SI{60}{\giga\electronvolt} < m_{\ell^+\ell^-} < \SI{120}{\giga\electronvolt}$ \\
  \bottomrule \\
\end{tabular}
\begin{tabular}{lccc}
  \toprule
  CMS \SI{8}{\tera\electronvolt}~\CMSCitation{} & \multicolumn{3}{c}{$t\bar{t}$ (68516.v1)} \\
  \midrule
  Observable & $\mathrm{d} \sigma / \mathrm{d} m_{t\bar{t}}$ (Table 39) & $\mathrm{d} \sigma / \mathrm{d} y_{t\bar{t}}$ (Table 36) & $\mathrm{d} \sigma / \mathrm{d} y_{t}$ (Table 21) \\
  \multirow{1}{*}{Cuts} & no cuts & no cuts & no cuts \\
  \bottomrule
\end{tabular}
\begin{tabular}{l|c|c}
\end{tabular}
\caption{Observables and corresponding cuts for all datasets discussed in the validation and application sections.
Each dataset name includes the \texttt{HepData} identifier, which can be used to download the measured cross sections.}
\label{tab:datasets}
\end{table*}

\begin{figure*}
  \centering
  \includegraphics[height=0.31\textheight]{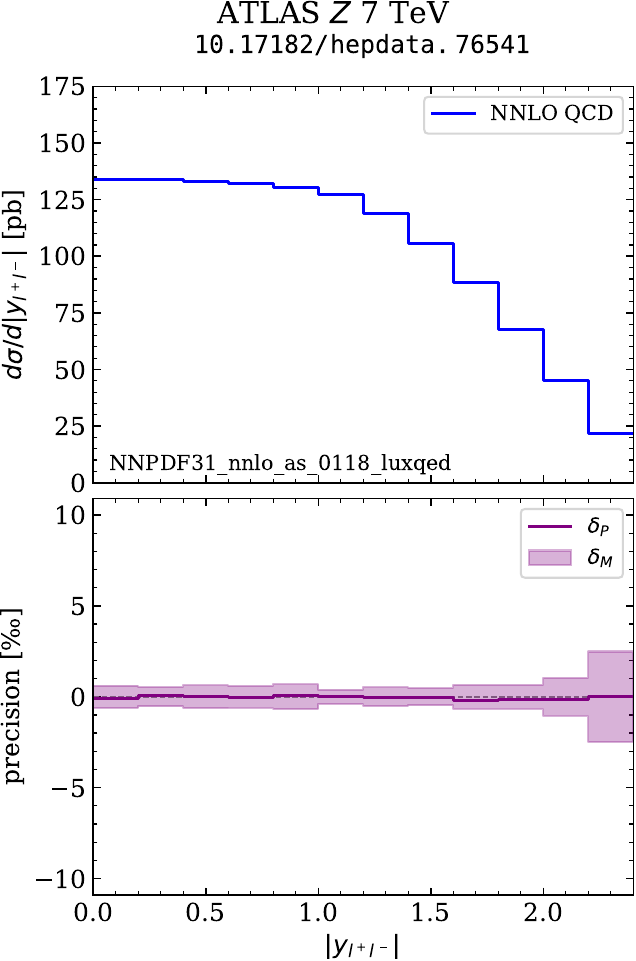}\hfill
  \includegraphics[height=0.31\textheight]{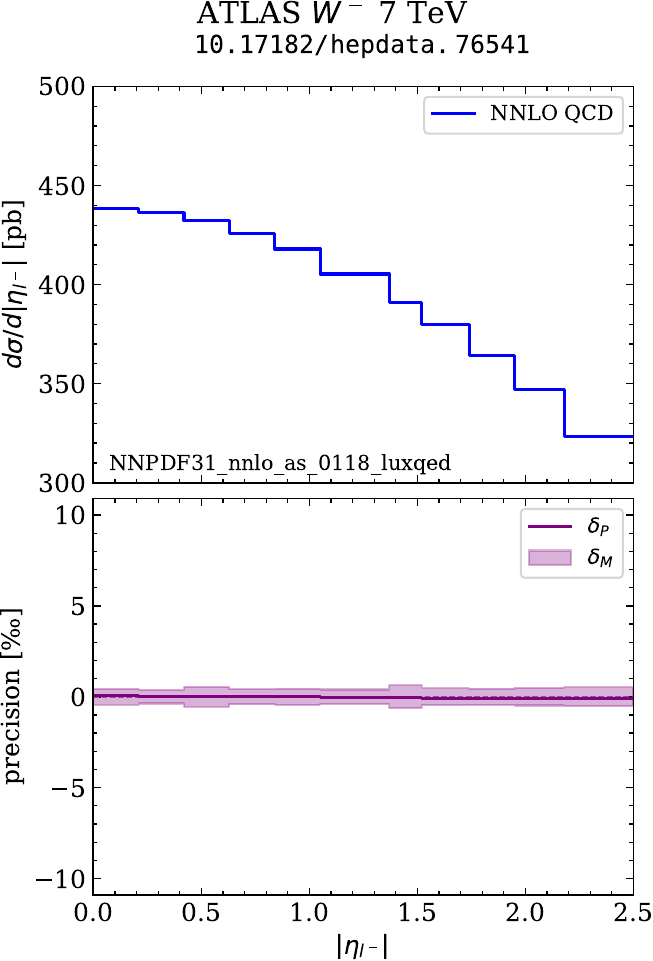}\hfill
  \includegraphics[height=0.31\textheight]{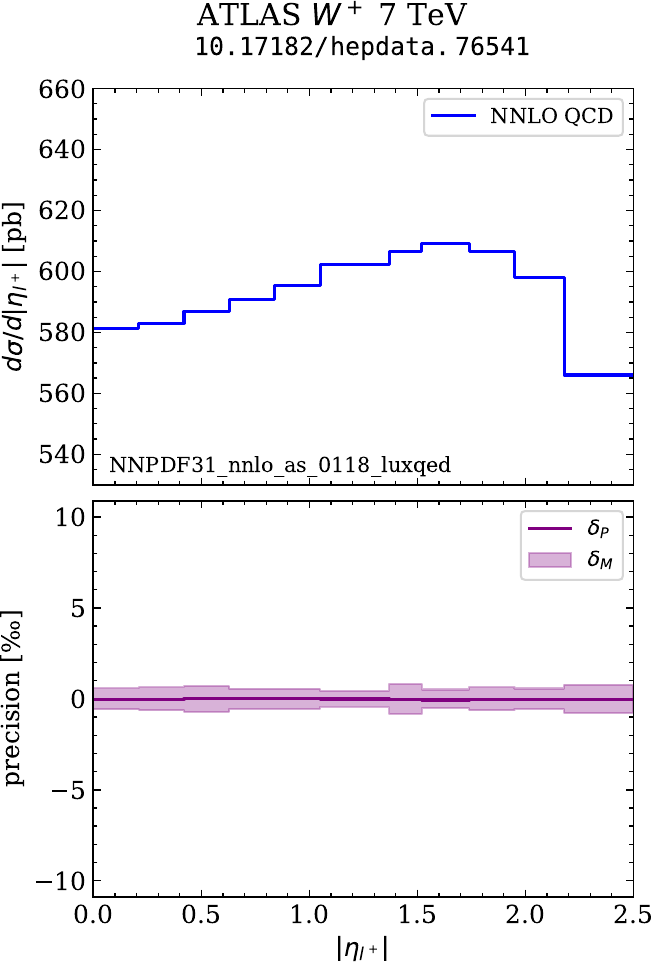}
  \caption{Comparison of \Matrix{} and \MatrixHawaii{} predictions for the (pseudo)rapidities of the lepton (pair) in NC (CC) in DY measured in ATLAS at 7 TeV~\ATLASCitation{}. 
  The lower panel shows the relative difference of the predictions as well as the uncertainties from \Matrix{}, which combine integration and \rcut{} extrapolation uncertainties.}
  \label{fig:validation_interpolation_ATLAS}
\end{figure*}

\begin{figure*}
  \centering
  \includegraphics[height=0.31\textheight]{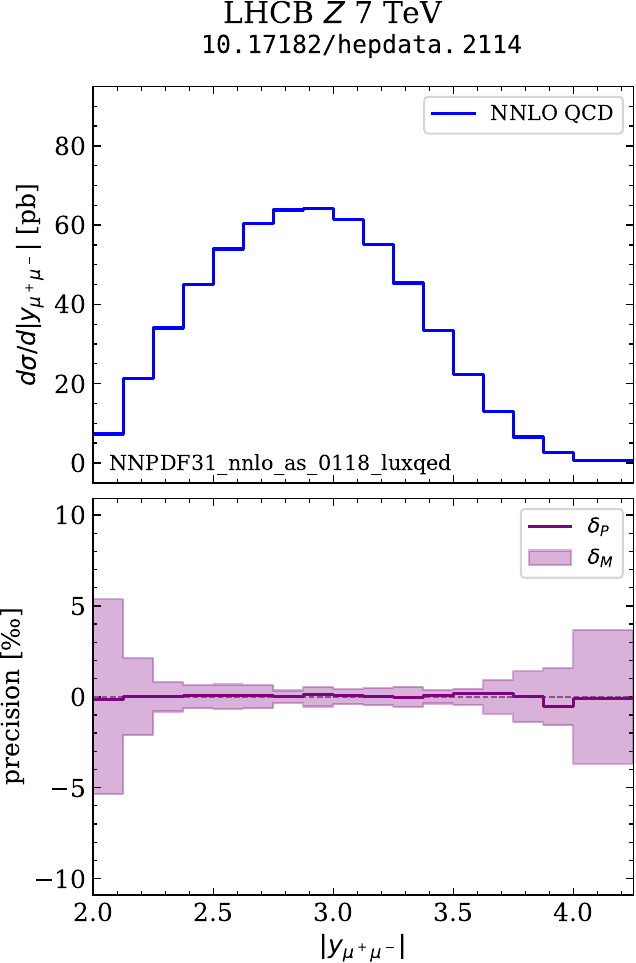}\hfill
  \includegraphics[height=0.31\textheight]{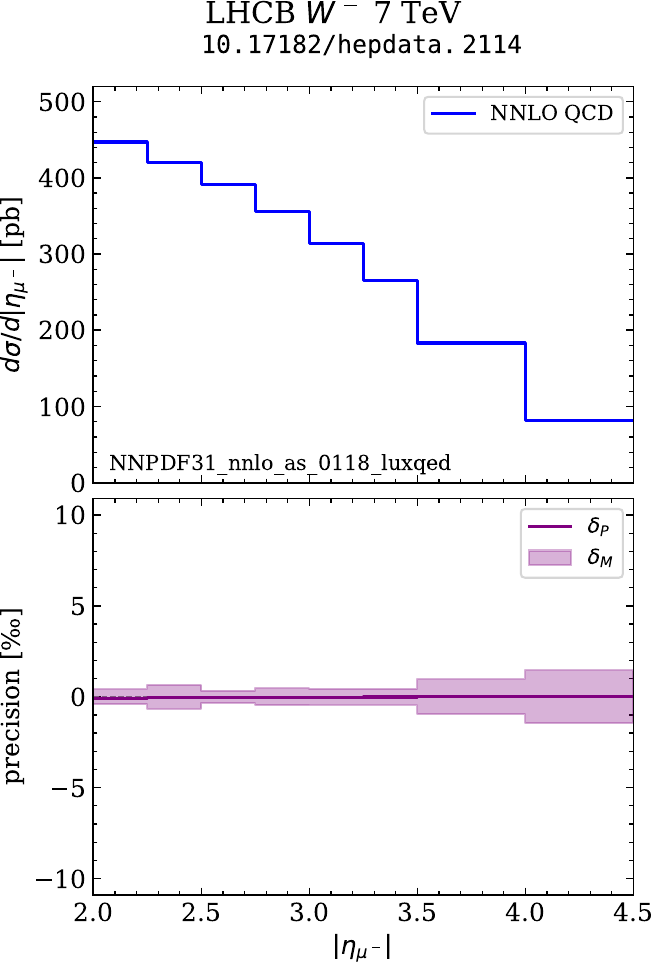}\hfill
  \includegraphics[height=0.31\textheight]{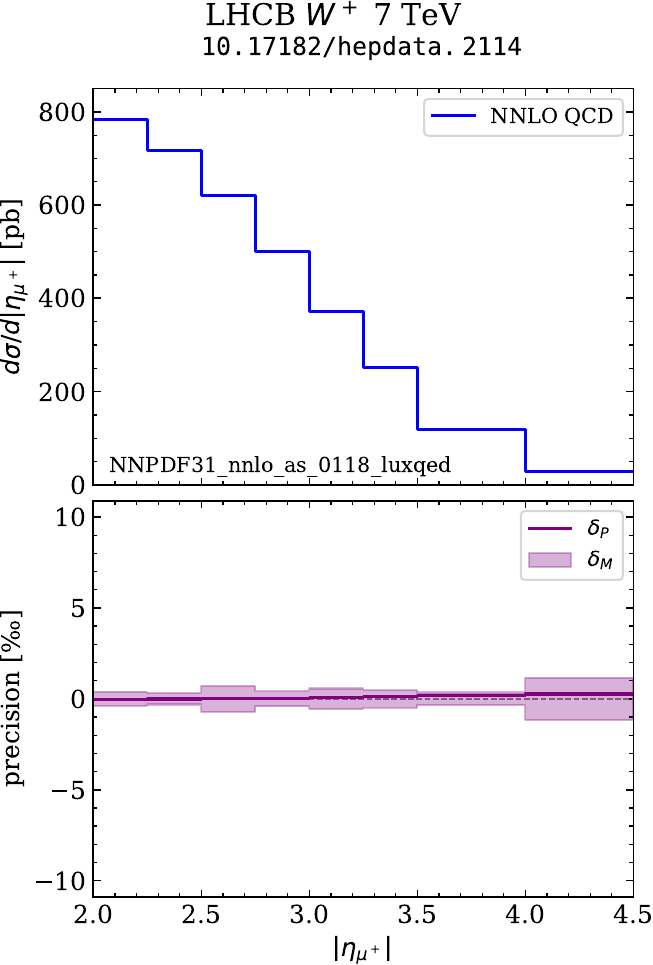}
  \caption{Comparison of \Matrix{} and \MatrixHawaii{} predictions for the (pseudo)rapidities of the lepton (pair) in NC (CC) in DY measured in LHCb at 7 TeV~\LHCbCitation{}. 
  The lower panel shows the relative difference of the predictions as well as the uncertainties from \Matrix{}, which combine integration and \rcut{} extrapolation uncertainties.}
  \label{fig:validation_interpolation_LHCB}
\end{figure*}

\begin{figure*}
  \centering
  \includegraphics[height=0.31\textheight]{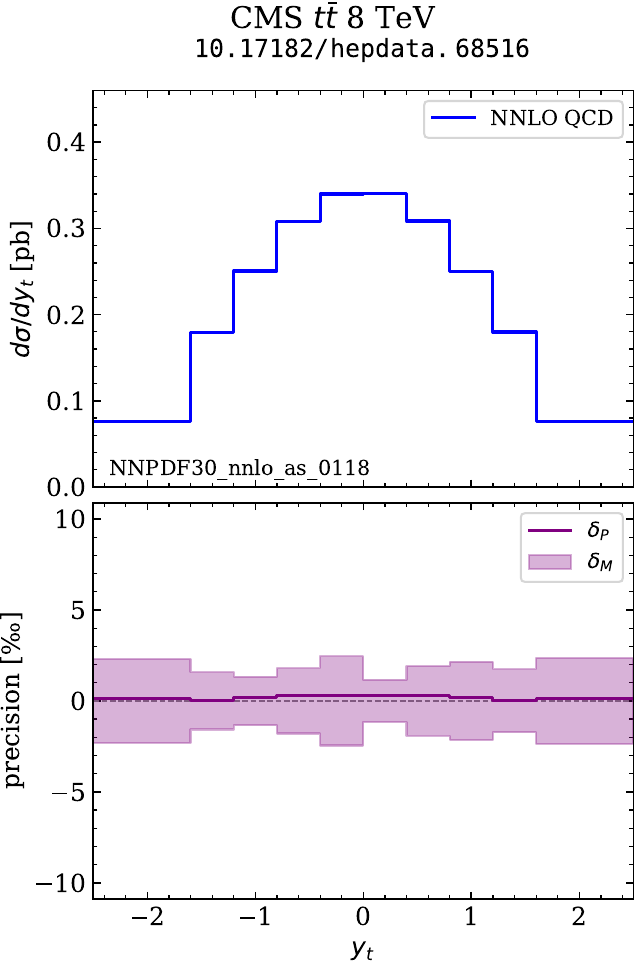}\hfill
  \includegraphics[height=0.31\textheight]{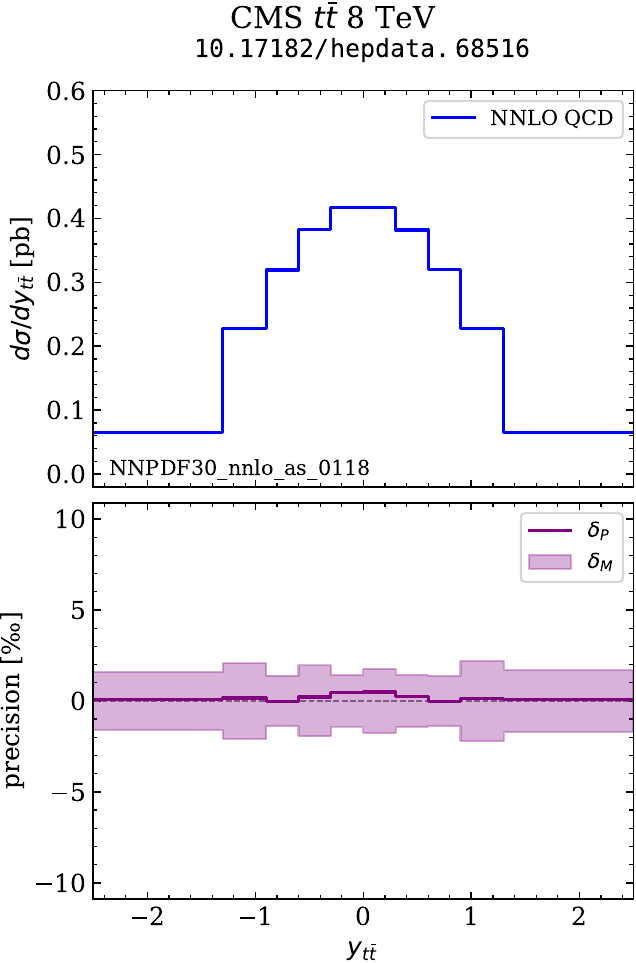}\hfill
  \includegraphics[height=0.31\textheight]{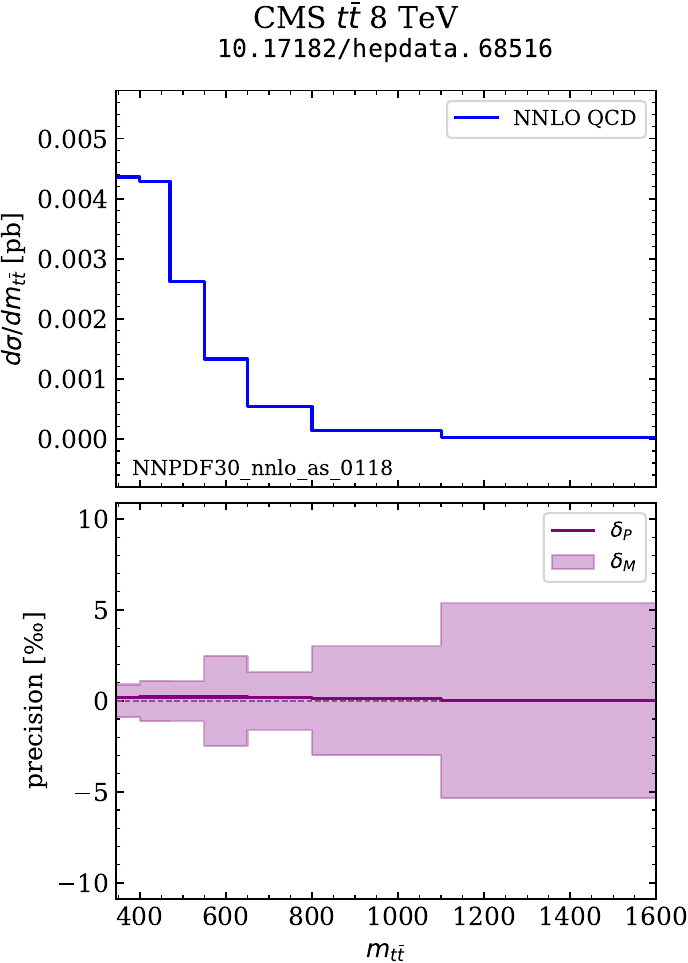}
  \caption{Comparison of \Matrix{} and \MatrixHawaii{} predictions for the rapidities of the top and the top-pair, and the transverse mass of the top-pair in top pair production measured in CMS at 8 TeV~\CMSCitation{}. 
  The lower panel shows the relative difference of the predictions as well as the uncertainties from \Matrix{}, which combine integration and \rcut{} extrapolation uncertainties.}
  \label{fig:validation_interpolation_CMS}
\end{figure*}

\subsection{Interpolation errors}

We start by assessing the relative interpolation error $\delta_\mathrm{P}$, defined as the difference between the result from \Matrix{} and the convolution of the interpolation grid with the same PDF set, relative to the \Matrix{} result:
\begin{equation}
\delta_\mathrm{P} = \frac{\text{\PineAPPL} - \text{\Matrix}}{\text{\Matrix}} \,.\\
\end{equation}
This is an appropriate measure of the interpolation error because the \PineAPPL{} grids and \Matrix{} results are constructed using
identical events and would agree perfectly, had the momentum fractions and scales not been interpolated in \PineAPPL{}.
To put this number into perspective, we compare it against the combined uncertainty from \Matrix{}, denoted as $\delta_M$, which covers both the MC (integration) uncertainty and the \rcut{} extrapolation uncertainty for NNLO QCD (see \Cref{sec:rcut-parameter-dependence}).

In \Cref{fig:validation_interpolation_ATLAS,fig:validation_interpolation_LHCB,fig:validation_interpolation_CMS} we show the absolute predictions, interpolation errors and combined \Matrix{} uncertainties for the \ATLAS{}, \LHCb{} and \CMS{} measurements.
In all cases we first observe that the \Matrix{} and \MatrixHawaii{} predictions agree very well and that the interpolation errors are below the per mille level and negligible w.r.t.\ the combined uncertainties from \Matrix{}, which are at the per mille level for most of the bins and always remain below half a percent.
Also note that the combined \Matrix{} uncertainties are much smaller than the NNLO 7-pt scale variation bands (not shown on these plots) and will thus be omitted in all following plots.

\subsection[$r_{\mathrm{cut}}$-parameter dependence]{\boldmath$r_{\mathrm{cut}}$-parameter dependence\label{sec:rcut-parameter-dependence}}

\begin{figure*}
  \centering
  \includegraphics[height=0.31\textheight]{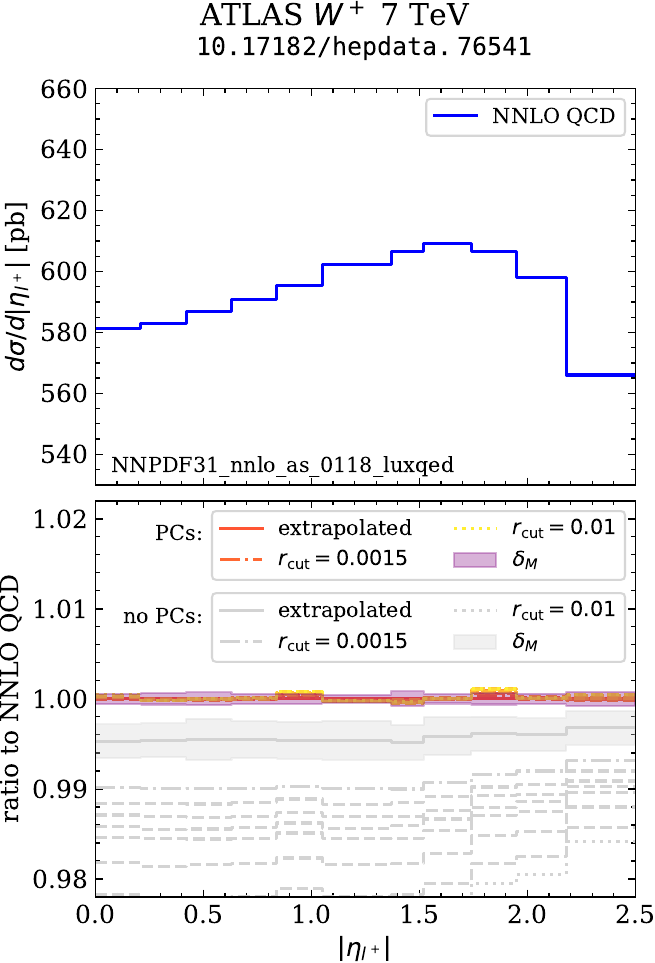}\hfill
  \includegraphics[height=0.31\textheight]{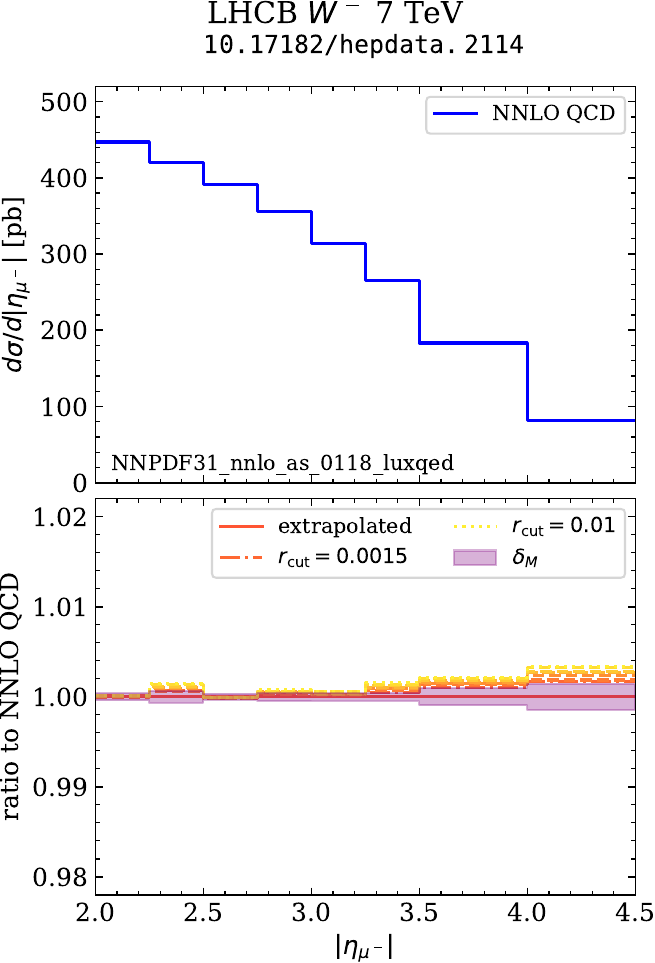}\hfill
  \includegraphics[height=0.31\textheight]{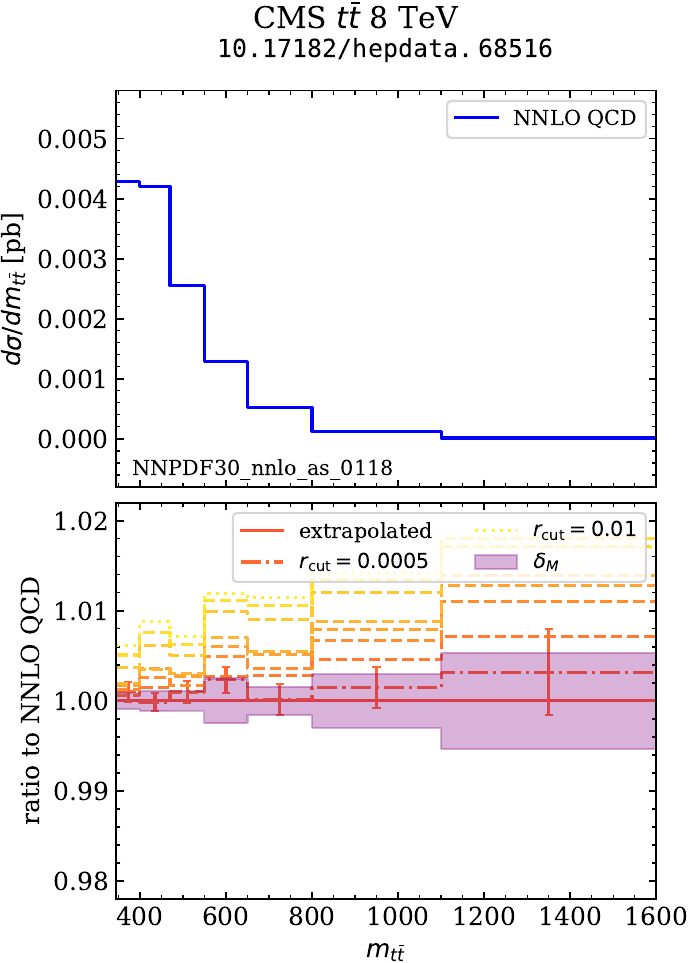}
  \caption{NNLO QCD predictions for ATLAS $W^+$ (left), LHCb $W^-$ (center) and top-quark pair (right) production.
    The ratio panels show predictions for eight different values of the \rcut{} parameter relative to the \mbox{$\rcut \to 0$} extrapolation, from the largest value of \rcut{} in yellow to the lowest in red.
    For ATLAS $W^+$ production (left) we show the results with and without (in gray) inclusion of linear fiducial power corrections,
    for top-quark pair production (right) we include the statistical error at the lowest value $\rcut=0.0005$ for reference.
  \label{fig:extrapolation}}
\end{figure*}

At NNLO, \Matrix{} relies on $q_\mathrm{T}$-subtraction, which uses the transverse momentum $q_\mathrm{T}$ of the final-state colour-singlet (plus heavy quarks) system to slice off phase space regions with \mbox{$\rcut < q_\mathrm{T} / M$}, where $M$ is the invariant mass of that system, which then are described by a counterterm derived from $q_\mathrm{T}$-resummation.
Thus the NNLO corrections pick up a dependence on this slicing parameter \rcut{}, which is removed by numerically extrapolating in \mbox{$\rcut \to 0$}. To be able to perform the extrapolation, in \Matrix{} the NNLO corrections are simultaneously calculated for a set of \rcut{} values, and a weighted least-squares fit with a quadratic polynomial as functional form is eventually applied on a subset of these $\rcut$ values.
This procedure is applied separately for each distribution on a bin-by-bin basis.
The numerical uncertainties quoted by \Matrix{} encapsulate both uncertainties from the Monte Carlo integration and from the fit at the core of the $\rcut\to0$ extrapolation.%
\footnote{ More details on the \mbox{$\rcut \to 0$} extrapolation, and the estimation of its uncertainty, can be found in Section 6 of Ref.\cite{Grazzini:2017mhc} for integrated cross sections. For distributions, such extrapolation has been publicly available since \Matrix{} {\tt v2.1}.}

In \MatrixHawaii{} this functionality is preserved: the interpolation grid provided to the user contains the extrapolated result in each bin, matching that of the corresponding distribution calculated directly by \Matrix{} below the per mille level, as shown in the previous Section.
To make this possible, we have generalized the extrapolation procedure to interpolation grids, based on intermediate interpolation grids automatically generated by our interface for the same set of \rcut{} values.
We note that the statistical errors at fixed $\rcut$ values, which are required to perform the fit, rely on the direct \Matrix{} results, since the \PineAPPL{} grids do not contain such error information themselves.
Some details on the extrapolation procedure for the interpolation grids are given in \Cref{app:extrapolation}.

In \Cref{fig:extrapolation} we show predictions for a subset of observables from the previously discussed set of measurements from
Refs.~\cite{ATLAS:2016nqi,LHCb:2015okr,CMS:2015rld}.
In the upper panels we present the differential distribution as obtained from the extrapolated interpolation grid, and in the lower panels we show a ratio of predictions obtained for the respective set of eight fixed $\rcut$ values spanning from \mbox{$\rcutmin = 0.0015$} and \mbox{$\rcutmin = 0.0005$} to \mbox{$\rcutmax = 0.01$} for the Drell--Yan and top-pair production processes, respectively.%
\footnote{More explicitly, this set of $\rcut$ values contains $0.0015$, $0.002$, $0.0025$, $0.003$, $0.0035$, $0.005$, $0.0075$, and $0.01$ for \mbox{$\rcutmin = 0.0015$}; and $0.0005$, $0.001$, $0.0015$, $0.002$, $0.0025$, $0.005$, $0.0075$, and $0.01$ for \mbox{$\rcutmin = 0.0005$}, respectively. 
This selection of $\rcut$ values corresponds to the \Matrix{} default for distributions and can be modified by the user through the standard \Matrix{} parameter {\tt switch\_accuracy}.}
The error bands shown refer to the aforementioned numerical uncertainties provided by \Matrix{}, including both extrapolation uncertainties and the statistical uncertainties of the fixed \rcut{} calculations, which are propagated through the extrapolation procedure.
Such statistical uncertainties might not be negligible, as shown in the right plot: they are reported only for the lowest value, where they are largest, to improve readability.
We note that, in the weighted fit, \rcut{} values with comparably large integration uncertainties only mildly affect the result.

The higher minimal value of $\rcut$ in the Drell--Yan process is counterbalanced by the inclusion of linear fiducial power corrections in $\rcut$~\cite{Buonocore:2021tke} for both considered setups for $Z$ production and for $W$ production in the ATLAS setup~\ATLASCitation, which leads to a significant reduction of the $\rcut$ dependence in regions where such power corrections arise, and thus to an improvement of the numerical convergence (left plot). 
For illustration, we also show the $\rcut$ dependence \textit{without} including the fiducial power corrections.%
\footnote{The purpose of showing the result also without inclusion of fiducial power corrections is twofold: first to illustrate that in this case the extrapolation is essential to get a reasonable result, second to confirm the \Matrix{} recommendation to refer to smaller $\rcut$ starting values to achieve reliable predictions for observables that receive significant power corrections in $\rcut$, unless those are explicitly corrected for.}
For $W$ production in the considered LHCb setup~\LHCbCitation{} linear fiducial power corrections are absent (central plot); the visible $\rcut$ dependence for large pseudorapidities is thus of different origin and controlled by the bin-wise extrapolation procedure.

The same holds true for the power corrections in inclusive top-pair production (right plot) in the CMS setup~\CMSCitation.
We observe that using a fixed, even quite low $\rcut$ value in several cases would result in predictions that differ significantly, up to the percent level, from the extrapolated results, showing the relevance of a bin-wise extrapolation.

\section{Applications of interpolation grids\label{sec:applications}}

The potential applications for the interpolation grids are vast.
In this section we showcase but a few of the possible applications of the interpolation grids that can be generated with \MatrixHawaii{}.
First, in \cref{sec:interpolation-grids-at-nnlo-qcd} we present NNLO QCD predictions for the aforementioned measurements by the ATLAS, CMS and LHCb collaborations at the LHC in the form of interpolations grids that can be used to improve PDF determinations.
The exact NNLO predictions obtained from these interpolation grids are compared, in \cref{sec:improving-predictions-for-pdf-determinations}, with analogous results where the NNLO corrections are approximated via $K$-factors, exploring the limitations of $K$-factor--based predictions.
In \cref{sec:interpolation-grids-at-nnlo-qcd-nlo-ew} we showcase the possibility to use the \MatrixHawaii{} framework to obtain interpolation grids at NNLO QCD which also contain NLO EW corrections.
Finally, in \cref{sec:PDFunc} we illustrate the flexibility of interpolation grids that allows us to easily change the PDF sets used in the predictions and to calculate PDF uncertainties.

While the primary purpose of this paper is to present \MatrixHawaii{}, a secondary purpose is the publication of the interpolation grids used in the following subsections, which can be directly included in PDF determinations.
We upload these grids to \texttt{PloughShare}~\cite{PloughShare}, which is a public database of interpolation grids, so that the community can reuse our predictions without having to invest considerable computational resources.

\subsection{Interpolation grids at NNLO QCD\label{sec:interpolation-grids-at-nnlo-qcd}}

\paragraph{ATLAS \boldmath$\gamma^*/Z$ and \boldmath$W^{\pm}$ at 7 TeV}

In Ref.~\ATLASCitation{} ATLAS measures the $W^+$, $W^-$ and $Z$ production processes in \SI{4.8}{\per\femto\barn} of data and compares them to NNLO QCD+NLO EW predictions obtained with six different PDF sets: ABM12~\cite{Alekhin:2013nda}, CT14~\cite{Dulat:2015mca}, HERAPDF2.0~\cite{H1:2015ubc}, JR14~\cite{Jimenez-Delgado:2014twa}, MMHT2014~\cite{Harland-Lang:2014zoa} and NNPDF3.0~\cite{NNPDF:2014otw}, showing a good potential of this dataset for further constraints in the majority of the PDF fits.

In the same publication~\ATLASCitation{}, this dataset is used to determine proton PDFs together with other neutral-current and charged-current DY data from ATLAS and DIS data from HERA.
There the previous observation of an unexpectedly large strange-quark density in the proton~\cite{ATLAS:2012sjl} is confirmed.
The first global analyses of this data find it to be in slight tension with other data sets~\cite{Thorne:2017aoa,Hou:2019gfw,NNPDF:2017mvq}.
As a consequence, it is originally not or only partially included in the main PDF releases.

Most of recent analyses describe the data at NNLO QCD, with EW corrections included in some cases~\cite{Amoroso:2022eow}.
Thorough comparisons of available NNLO QCD predictions in Ref.~\cite{Alekhin:2021xcu,Alekhin:2024mrq} highlight that the data is now precise
enough to make na\"{i}ve slicing approaches insufficient if too large fixed slicing cut values are used. This is in particular due to the symmetric cuts applied in this analysis, which give
rise to fiducial linear power corrections. In our calculation we solve this issue
by explicitly including them~\cite{Buonocore:2021tke} and, moreover, numerically
extrapolate to the limit of a vanishing slicing cut, as discussed in the previous section.
Predictions at higher formal accuracy up to next-to-next-to-next-to-leading order (N3LO) QCD~\cite{Duhr:2020sdp,Camarda:2021ict,Chen:2021vtu,Duhr:2021vwj,Chen:2022cgv,Chen:2022lwc} and including resummation~\cite{Amoroso:2022lxw,Neumann:2022lft,Camarda:2023dqn} are also available.

\begin{figure*}[t]
  \centering
  \includegraphics[height=0.31\textheight]{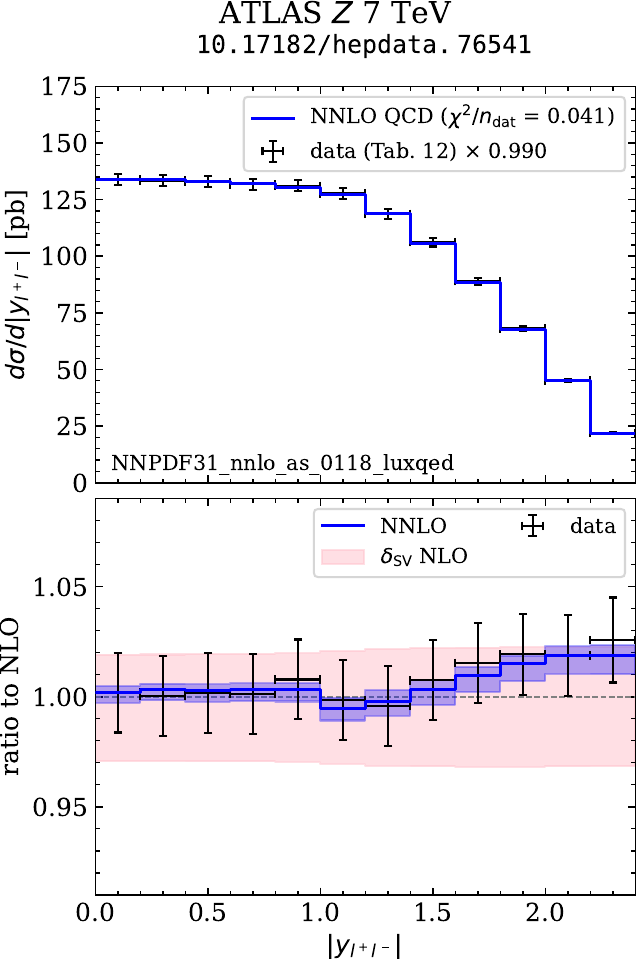}\hfill
  \includegraphics[height=0.31\textheight]{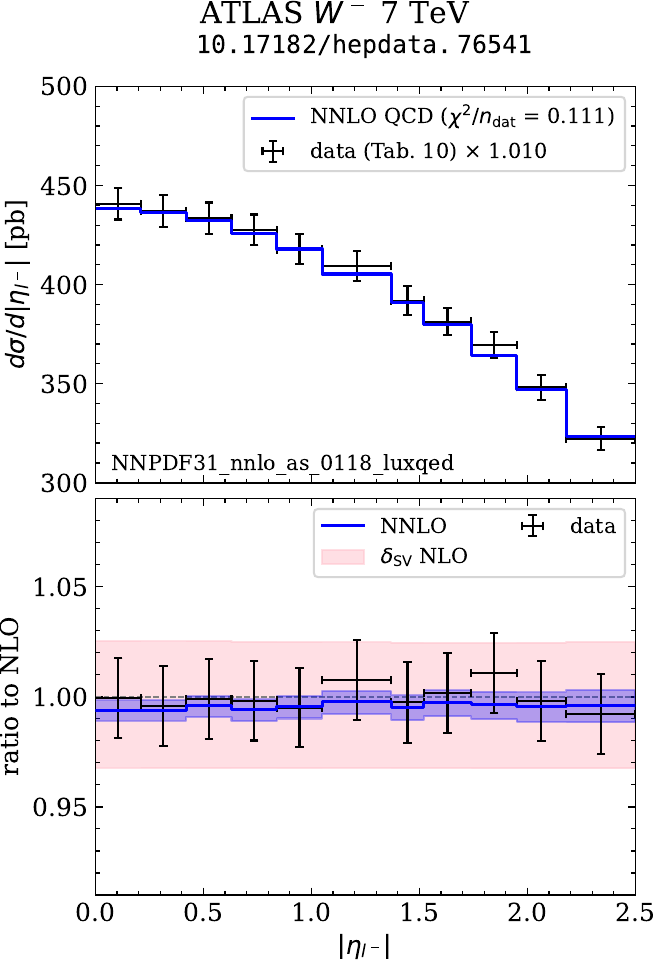}\hfill
  \includegraphics[height=0.31\textheight]{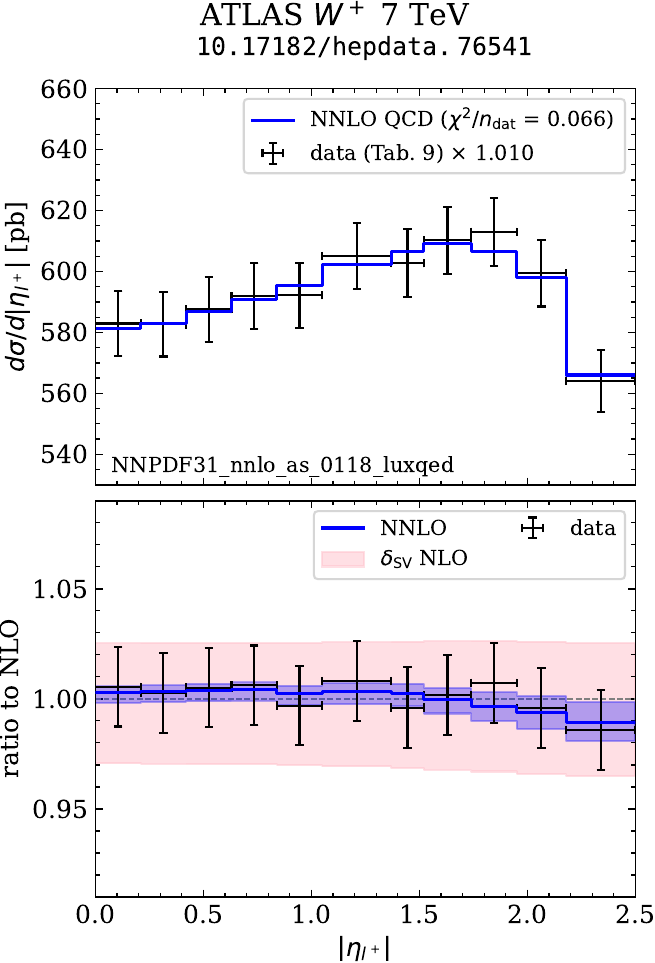}
  \caption{Comparison of \MatrixHawaii{} predictions to $Z$, $W^-$ and $W^+$ boson production data from ATLAS at 7
  TeV~\ATLASCitation{}.  The upper panel shows the NNLO QCD prediction in blue and data, {\tt HepData id:} 76541.v1, in black.
  Quality of data description, quantified via the {\em na\"ive} $\chi^2$ value per data point, is printed in the legend.  The data is
  normalized such as to minimize this $\chi^2$, for each production mode individually, and the fitted normalization factor is
  also printed in the legend. The lower panel shows the NNLO $K$-factor and the relative NLO and NNLO 7-pt scale variation (SV)
uncertainties.}
  \label{fig:atlas7tev_dataToTheory}
\end{figure*}

\Cref{fig:atlas7tev_dataToTheory} shows our predictions for a subset of measurements in this data set, on {\tt HepData} available under the id 76541, convolved with the {\tt NNPDF31\_nnlo\_as\_0118\_luxqed} PDF set~\cite{Bertone:2017bme}.
In particular, we compare to cross sections for \mbox{$Z \to l^+ l^-$} production in $|y_{l^+ l^-}|$ bins in the ``on-shell region'' with \mbox{$66\text{
GeV}<m_{l^+l^-}<116\text{ GeV}$} ({\tt HepData} Table 12) and the cross sections for \mbox{$W^+\to l^+\nu$} (Table 9) and \mbox{$W^-\to l^-\bar{\nu}$}
(Table 10) production in $\eta_{l^\pm}$ bins, combined from \mbox{$l \in \{ e,\mu\}$} decays and extrapolated to the common fiducial region.
Also measured, but not compared to here, are the cross sections for \mbox{$Z \to l^+ l^-$} production in two ``off-shell regions'', \mbox{$46\text{ GeV}<m_{l^+l^-}<66\text{ GeV}$} and \mbox{$116\text{ GeV}<m_{l^+l^-}<150\text{ GeV}$}.

The lower panels report the frequently used 7-pt scale variation uncertainties at NLO and NNLO QCD as pink and blue error bands,
respectively.
For each production mode the inclusion of higher-order corrections reduces the uncertainty band appreciably from a few percent down to about one percent.
The NNLO $K$-factors, plotted in blue, fall into the NLO scale variation band and are below one percent across most of the range.
The NNLO prediction describes the data extremely well, see the values of {\em na\"ive} $\chi^2$ per data point\footnote{We define {\em na\"ive} $\chi^2$ as $\sum_i^{n_{\text{dat}}} (T_i - D_i)^2 / \Delta_i^2$ where $T_i$ is the theoretical prediction, $D_i$ the measurement, $\Delta_i$ the measurement uncertainty, and $n_{\text{dat}}$ the number of data points. This definition intentionally omits theoretical uncertainties as traditional PDF determinations commonly do.} reported on the plot, which is reassuring since it is included in the  {\tt NNPDF31\_nnlo\_as\_0118\_luxqed} PDF fit.

The results presented in this section were obtained using the \PineAPPL{} interpolation grids published alongside this publication.
The numerical errors of the corresponding \Matrix{} run and the interpolation errors are discussed in \Cref{sec:validation} and are both sub-leading as compared to the NNLO scale variation uncertainties. 

\paragraph{LHCb \boldmath$\gamma^*/Z$ and \boldmath$W^{\pm}$ 7 TeV}

\begin{figure*}[t]
  \centering
  \includegraphics[height=0.31\textheight]{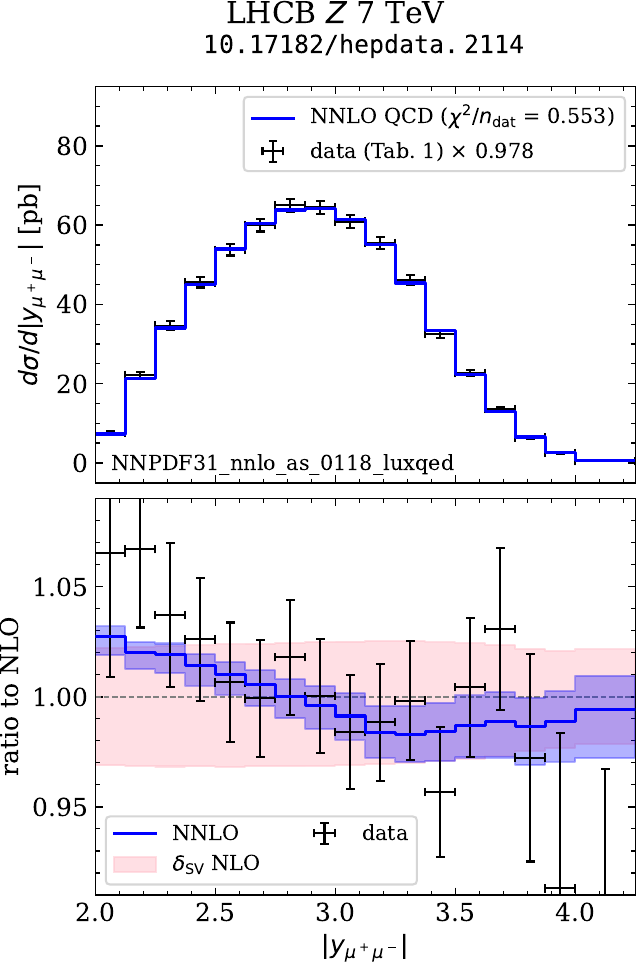}\hfill
  \includegraphics[height=0.31\textheight]{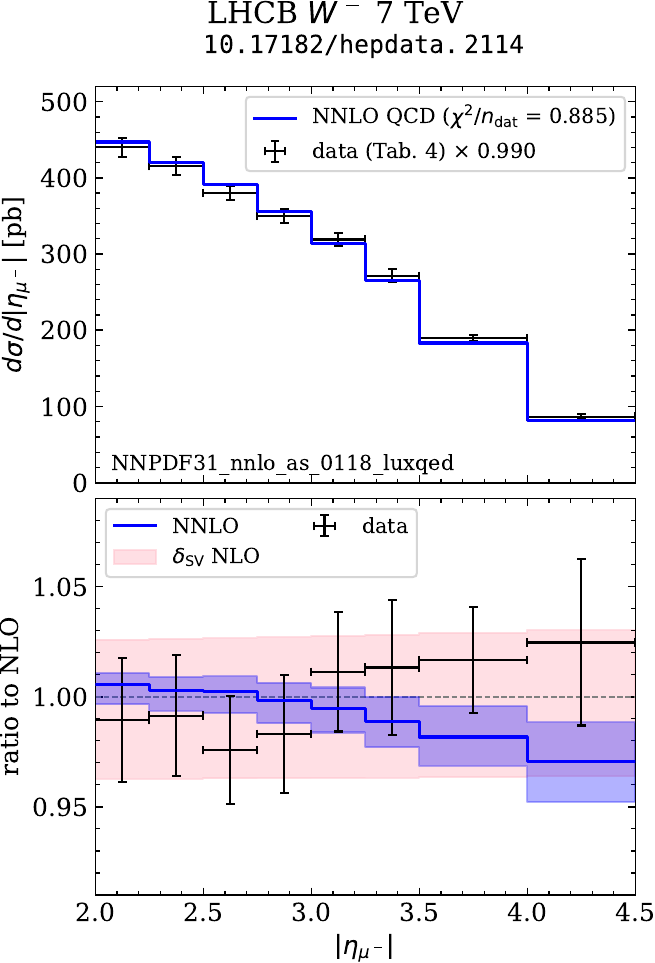}\hfill
  \includegraphics[height=0.31\textheight]{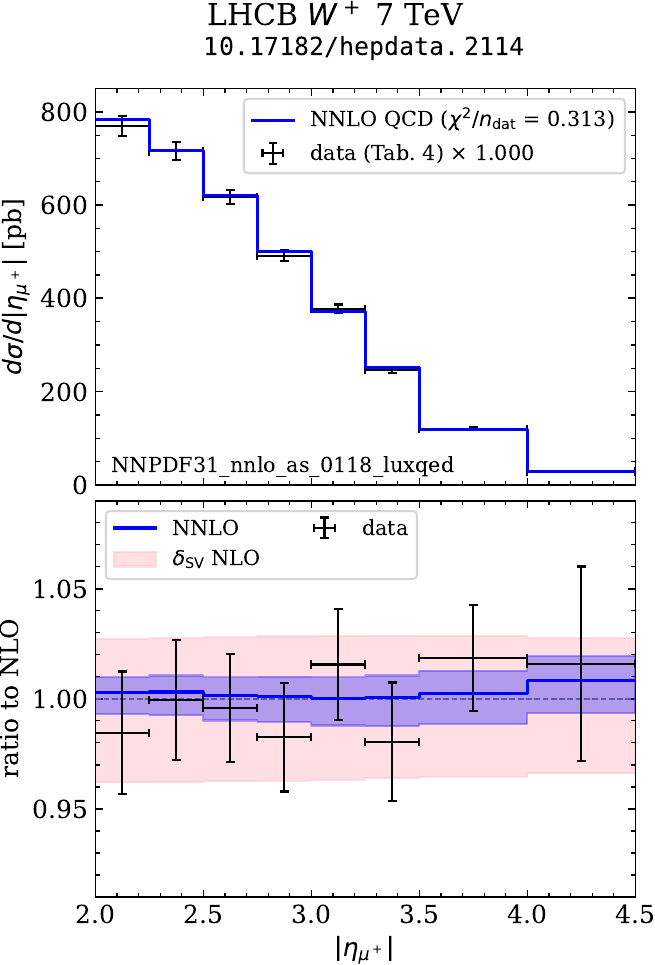}
  \caption{Comparison of \MatrixHawaii{} predictions to $Z$, $W^-$ and $W^+$ boson production data from LHCb at 7
  TeV~\LHCbCitation{}.  The upper panel shows the NNLO QCD prediction in blue and data, {\tt HepData id:} 2114.v1, in black.
  Quality of data description, quantified via the {\em na\"ive} $\chi^2$ value per data point, is printed in the legend.  The data is
  normalized such as to minimize this $\chi^2$, for each production mode individually, and the fitted normalization factor is
  also printed in the legend. The lower panel shows the NNLO $K$-factor and the relative NLO and NNLO 7-pt scale variation (SV)
  uncertainties.}
  \label{fig:lhcb7tev_dataToTheory}
\end{figure*}

The measurement of $Z$ and $W^\pm$ boson production at LHCb during the 7 TeV $pp$ run at the LHC run was presented in Ref.~\LHCbCitation{}. 
It provides cross sections, rapidity spectra and various ratios, including lepton charge asymmetries, with the integrated luminosity of 1 fb$^{-1}$.
For proton structure determinations, this measurement is complementary to those in general purpose detectors, like ATLAS or CMS, because of a different kinematic reach in rapidity spanning the range between 2 and 4.5.

The experimental data can be satisfactorily described only if NNLO QCD corrections are included~\cite{NNPDF:2017mvq} and, similar to the 7 TeV ATLAS data, it comes with measurement uncertainties smaller than the PDF uncertainties of the majority of the predictions obtained with the ABM12~\cite{Alekhin:2013nda}, CT10~\cite{Gao:2013xoa}, HERA15~\cite{Cooper-Sarkar:2010yul}, JR09~\cite{Jimenez-Delgado:2008orh}, MSTW08~\cite{Martin:2009iq} and NNPDF3.0~\cite{NNPDF:2014otw} PDF sets.
This lead to its swift inclusion in global analyses of collinear proton structure from the ABMP~\cite{Alekhin:2015cza, Alekhin:2017kpj,Alekhin:2023uqx}, CTEQ-TEA~\cite{Hobbs:2019gob,Hou:2019efy,Xie:2021equ,Yan:2022pzl,Hou:2022onq}, MSHT~\cite{Bailey:2020ooq,Cridge:2021pxm,McGowan:2022nag,Cridge:2023ryv,Cridge:2024exf,Harland-Lang:2024kvt}, NNPDF~\cite{NNPDF:2017mvq,Ball:2017otu,Bertone:2017bme,Ball:2018twp,NNPDF:2019ubu,Nocera:2019wyk,AbdulKhalek:2020jut, NNPDF:2021njg} and nCTEQ~\cite{Risse:2024kmr} groups as well as in various joint efforts~\cite{PDF4LHCWorkingGroup:2022cjn,Jing:2023isu}.
The flavour separation and the constraining power on the valence distributions in the $x>\si 0.1$ region were first to be highlighted~\cite{Rojo:2017xpe}, but it also has an appreciable impact on $\bar{d}$ and $\bar{u}$ distributions at low $x$~\cite{Hou:2019efy}.
Further, these data have been used to extract the strong coupling~\cite{dEnterria:2019aat,Cridge:2024exf}, as additional constraints in the determinations of nuclear effects in light nuclei~\cite{Alekhin:2017fpf}, and in simultaneous analyses of proton structure and SMEFT~\cite{Greljo:2021kvv,Kassabov:2023hbm}.
Moreover, the data has been very useful in proton structure determinations in multiple dimensions~\cite{Scimemi:2017etj, Bertone:2019nxa,Scimemi:2019cmh,Hautmann:2020cyp,Bury:2022czx,Moos:2023yfa,Bacchetta:2024qre,Bacchetta:2022awv}.

In \Cref{fig:lhcb7tev_dataToTheory} we show our \MatrixHawaii{} predictions at NNLO for this data set convolved with the {\tt NNPDF31\_nnlo\_as\_0118\_luxqed} PDF set.
These data are available on {\tt HepData} under id 2114 in Table 1 for the process $pp \to \mu^+ \mu^-$, and in Table 4 for the $pp \to \mu^+\nu_\mu$ and $pp \to \mu^-\nu_\mu$ processes.

As before, the upper panels show the NNLO predictions and the data, and the lower panels report NLO and NNLO QCD scale uncertainties.
The data are described very well, as expected, as it is part of the {\tt NNPDF31\_nnlo\_as\_0118\_luxqed} analysis.
The NNLO corrections reduce the scale uncertainties significantly, the NNLO $K$-factors reach up to 3 percent in the tails of the distributions, but are still contained in the NLO scale variation bands.

Finally, the \PineAPPL{} interpolation grids used for these predictions are published alongside this publication, c.f.~\Cref{sec:validation} for the discussion of the numerical and interpolation errors.

\paragraph{CMS \boldmath$t\bar{t}$ 8 TeV}
\begin{figure*}[t]
  \centering
  \includegraphics[height=0.31\textheight]{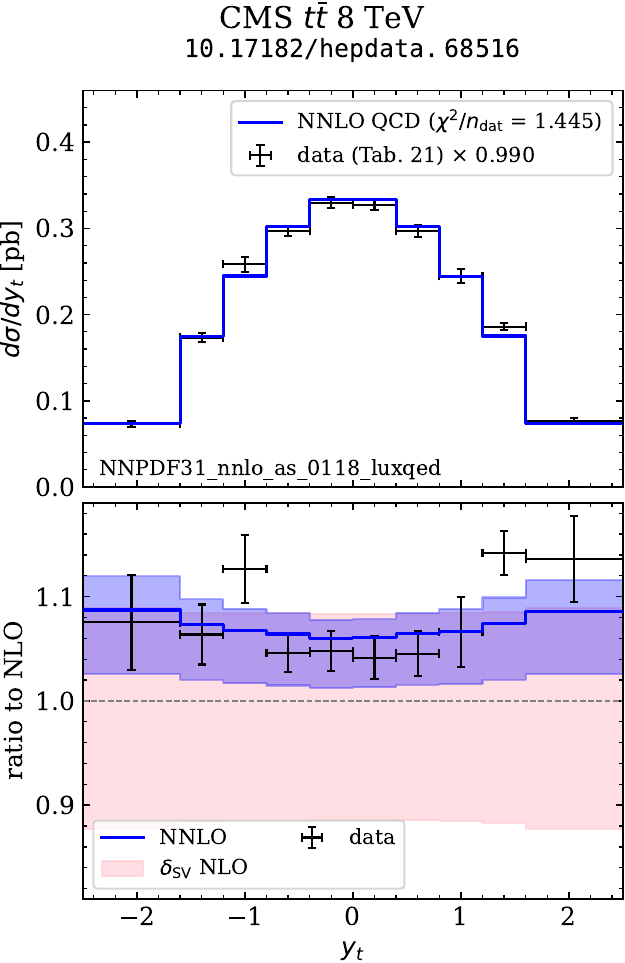}\hfill
  \includegraphics[height=0.31\textheight]{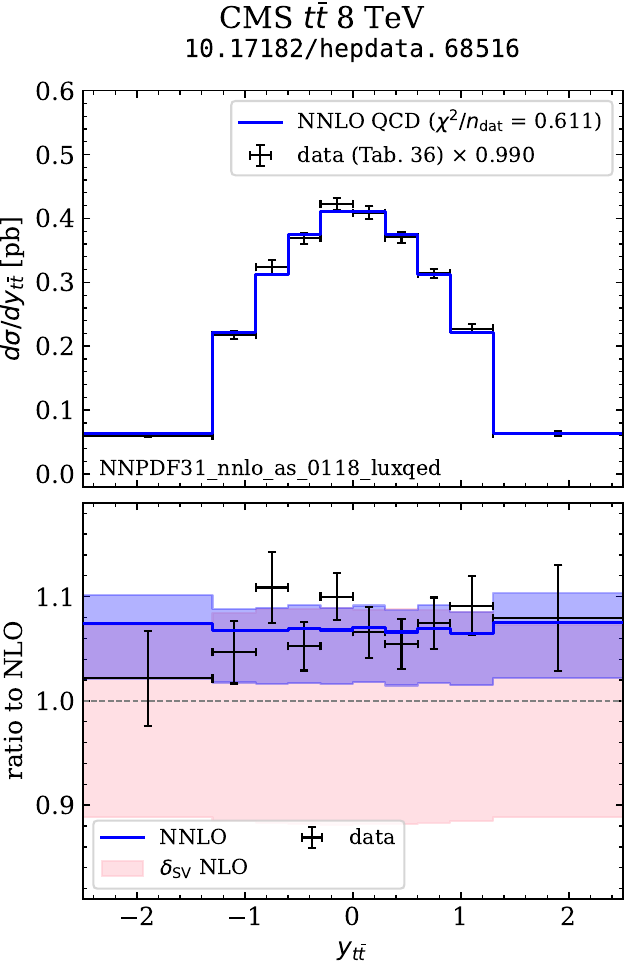}\hfill
  \includegraphics[height=0.31\textheight]{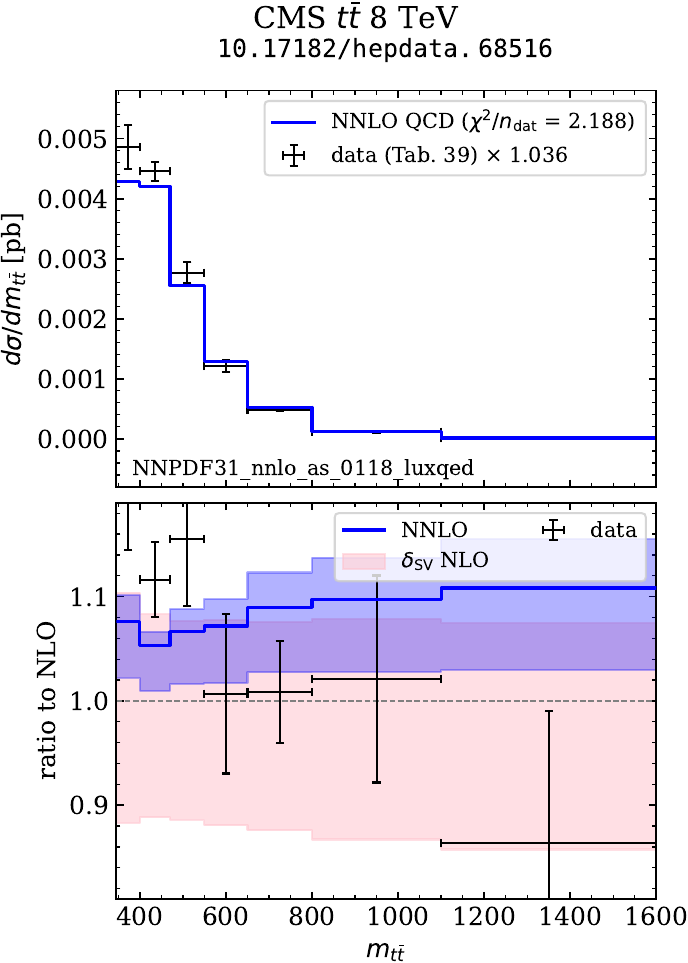}
  \caption{Comparison of \MatrixHawaii{} predictions to top pair production data from CMS at 8 TeV~\CMSCitation{}
  The upper panel shows the NNLO QCD prediction in blue and data, {\tt HepData id:} 68516.v1, in black.
  Quality of data description, quantified via the {\em na\"ive} $\chi^2$ value per data point, is printed in the legend.  The data is
  normalized such as to minimize this $\chi^2$, for each spectrum individually, and the fitted normalization factor is
  also printed in the legend. The lower panel shows the NNLO $K$-factor and the relative NLO and NNLO 7-pt scale variation (SV)
  uncertainties.}
  \label{fig:cms_dataToTheory}
\end{figure*}
In Ref.~\cite{CMS:2015rld} the CMS collaboration presented their measurement of top-quark pair production cross section, extracted from 19.7 fb$^{-1}$ of data relative to the \SI{8}{\tera\electronvolt} run.
The top quarks were reconstructed from the lepton+jet ($e^\pm/\mu^\pm$+jet) and dilepton ($e^+ e^-$, $\mu^+ \mu^-$, $e^\pm \mu^\mp$) decay channels, and results were presented for multiple differential distributions, including transverse momenta, rapidities, and invariant masses of the leptons, $b$ jets, top quarks and the $t\bar t$ system. 
The various observables were then compared with different SM predictions, obtained by \textsc{MadGraph}(LO)~\cite{Alwall:2011uj}+\textsc{Pythia}~\cite{Sjostrand:2006za}, \textsc{Powheg}(NLO)~\cite{Alioli:2009je, Alioli:2010xd, Re:2010bp}+\textsc{Pythia}, \textsc{Herwig}~\cite{Corcella:2000bw}, MC@NLO~\cite{Frixione:2002ik}+\textsc{Herwig}, as well as further computations including resummation effects (NLO+NLL)~\cite{Ferroglia:2013zwa, Li:2013mia}, and approximate NNLO corrections~\cite{Kidonakis:2012rm}.

Successive theoretical studies have shown that the inclusion of higher orders in the perturbative expansion significantly improves the agreement between theory and experiment. 
Ref.~\cite{Czakon:2015owf} provided exact NNLO QCD corrections for the distributions measured by CMS which, together with the further inclusion of NLO EW corrections~\cite{Pagani:2016caq, Campbell:2016dks}, proved itself of primary importance for an accurate description of the data.
Resummation effects have also been included in Ref.~\cite{Pecjak:2016nee}, where they have been shown to play a major role in the boosted-top region.

After the computation of first NLO interpolation grids for this dataset~\cite{Czakon:2017dip, Carrazza:2020gss}, it became one of the standard measurements entering several PDF fits, used by the NNPDF~\cite{NNPDF:2017mvq, Bertone:2017bme, NNPDF:2021njg, NNPDF:2019ubu, Ball:2017otu} and MSHT~\cite{Bailey:2019yze, Bailey:2020ooq, McGowan:2022nag, Cridge:2023ryv} collaborations.
Furthermore, it is a staple measurement used in the global fits of SMEFT parameters~\cite{Ellis:2020unq, Ethier:2021bye, Hartland:2019bjb, Buckley:2015lku, Brivio:2019ius, Bartocci:2023nvp, Elmer:2023wtr}, also including simultaneous fits of PDFs and SMEFT parameters~\cite{Gao:2022srd,Kassabov:2023hbm}.
Finally, the same distributions have also been used in several different precision studies, such as the extraction of the strong coupling at approximate N3LO~\cite{Cridge:2024exf}, the study of its sensitivity to the gluon PDF at large $x$~\cite{Czakon:2016olj} and the constraints it gives on the top mass and the strong coupling~\cite{Cridge:2023ztj,Garzelli:2023rvx}. 

Our NNLO predictions obtained with \MatrixHawaii{} for this dataset are shown in~\Cref{fig:cms_dataToTheory}, convolved with the {\tt NNPDF31\_nnlo\_as\_0118\_luxqed} pdf set. 
As for the plots previously presented, in the upper panel we show the comparison between the NNLO prediction and the data, while in the lower panel we focus on the  scale variations. 
The data can be obtained on {\tt HepData} under id 68516.v1.
The good agreement between theory and experiment is, also in this case, expected, since the data set enters into the fit of the PDF set used in our prediction. 
A notable exception is the small invariant-mass region, where the theory prediction undershoots the experimental data, as already noted in several studies~(see e.g.~Refs.~\cite{CMS:2018htd, CMS:2018adi, ATLAS:2019hau, ATLAS:2020ccu}). 
We can further observe a significant reduction of the scale variation band, and thus of the perturbative uncertainties, with the inclusion of the NNLO corrections.

As for the other predictions presented in this paper, the corresponding \PineAPPL{} grids are published alongside this publication. 
Their numerical and interpolation errors are discussed in detail in~\Cref{sec:validation}.

\subsection[PDF determinations with $K$-factor--based vs.~exact NNLO calculations]{PDF determinations with \boldmath$K$-factor--based vs.~exact NNLO calculations\label{sec:improving-predictions-for-pdf-determinations}}

PDFs are a crucial component in the theoretical predictions of cross sections at hadron colliders.
They encapsulate our knowledge of the collinear-momentum distribution of partons within the proton, and their accurate determination is essential for precision tests of the SM as well as searches for new physics.
In a PDF determination study, \cref{eq:factorization-formula} is essentially inverted, so that PDFs $f_a^\mathrm{p} (x_1, Q^2)$ are fit from measurements $\sigma$ and their corresponding predictions $\sigma_{ab}^{(n,m)} (x_1, x_2, Q^2)$.
This requires repeated computations of NNLO predictions with different PDF parameters, a very computationally expensive procedure because of the time and resources necessary for accurate higher-order computations. 
This makes interpolation grids a pivotal asset for the PDF determination programme.

Due to the limited availability of NNLO QCD interpolation grids, for LHC measurements often only NNLO K-factors are used.
This approximation has several possible shortcomings.
The $K$-factors, while tailored to each measurement and derived bin-by-bin, are assumed to be PDF independent, which allows them to be evaluated once and reused across multiple fits, even by different collaborations.
Moreover, the $K$-factors rescale all production channels, which depend on different PDFs, uniformly, despite the fact that the NNLO corrections to different channels can vary significantly.

So far it is unknown what the impact of this approximation is, and a definitive answer can only be given by comparing several global PDF fits performed with $K$-factors and with exact NNLO QCD predictions.
While such a detailed study goes beyond the scope of this paper, in the following we explore the limitations of the traditional $K$-factor approach by comparing it to a calculation based on exact NNLO QCD predictions.
To that end, we incorporate the datasets already discussed in \cref{sec:validation} and listed in \cref{tab:datasets} into a test PDF analysis and examine the effect of using $K$-factor--approximated NNLO predictions on the quality of the data description.
Both predictions are obtained by convolving the \PineAPPL{} interpolation grids, with one important difference: the exact prediction explicitly depends on the PDF throughout, whereas this dependence is intentionally dropped in the NNLO $K$-factor of the $K$-factor--based approach.\footnote{Thus in the $K$-factor--based approach the theoretical prediction is obtained by convolving NLO interpolation grid multiplied by bin-by-bin NNLO/NLO $K$-factors that are kept fixed and not updated as the PDF changes.}

As the test PDF set we chose the CJ22 PDF set~\cite{Accardi:2023gyr}.
Our choice intends to minimise bias, in that we selected a recent PDF analysis that does not include any of the data analysed here and yet describes it relatively well.
However, the outcome of a comparison, like the one attempted in this study, must depend on the test PDF set by construction, and so this section should be viewed as a case study of instances where the use of $K$-factors may be inadequate. 

To emulate conditions in a PDF fit, we do not just inspect predictions convolved with CJ22 PDFs, but also generate variations by
modifying the values of a subset of the parameters of the CJ22 functional form at the input scale, one at a time.
We then plot the figure of merit, capturing the quality of the data description, as a function of those parameter values, both for
the exact predictions and for those relying on $K$-factors, and compare them.
As data we consider all three measurements (nine observables) discussed in the previous sections: NC and CC DY in \ATLAS{} and
\LHCb{} at 7 TeV, and top-pair production in CMS at 8 TeV~\CMSCitation{}. 
We consider three spectra from the CMS top pair measurement instead of just one.
In a regular PDF fit this would lead to ``double counting'', in that the constraints on proton structure from different spectra in the same measurement
overlap.
However, this is of no consequence here because it only changes the relative weight of this dataset, which enters identically in the exact NNLO and K-factor–based treatments.
The number of data points adds up to a total of 94.
As the figure of merit, we use the {\em na\"ive} $\chi^2$, $\chi^2 = \sum_i^{n_{\text{dat}}} (T_i - D_i)^2 / \Delta_i^2$\,, where $T_i$ is our
theoretical prediction, $D_i$ the measurement, and $\Delta_i$ the measurement uncertainty.\footnote{{The {\em na\"ive} prescription for the calculation of $\chi^2$ omits correlated uncertainties. This can have the impact of underestimating the absolute value of~$\chi^2$ and in turn influence how strongly each individual data set contributes to the total $\chi^2$. However, our main quantity of interest is the ratio of the ``K-factor'' to the ``exact NNLO'' $\chi^2$. For this ratio, the impact of correlated systematics is expected to largely cancel, because for both the $\chi^2$ values are computed for the same data and with the same experimental uncertainties. Moreover, only one of the data sets considered in this study reports correlated uncertainties.}} 
We adjust the normalization of the measurements, $N$, for each experiment individually by minimizing the $\chi^2$ in the CJ22 fit,
to $N_{\text{ATLAS}} = 0.951$, $N_{\text{LHCb}} = 0.977$, and $N_{\text{CMS}} = 1.014$, yielding a $\chi^2/n_{\text{dat}} \sim 1.4$. %127/94
The $K$-factors in the $K$-factor--based predictions are evaluated once, for one set of values of the PDF parameters, and are then kept fixed.
More details regarding this are discussed below.
This setup allows us not only to evaluate the PDF dependence of the $K$-factors but also to assess their impact on measurements
including their uncertainties.
Furthermore, we consider a range of measurements across a selection of processes, allowing us to implicitly take into account
issues like the relative constraining power of a specific dataset and the compatibility between various datasets.

\begin{figure*}
    \centering
    \includegraphics[width=0.48\textwidth]{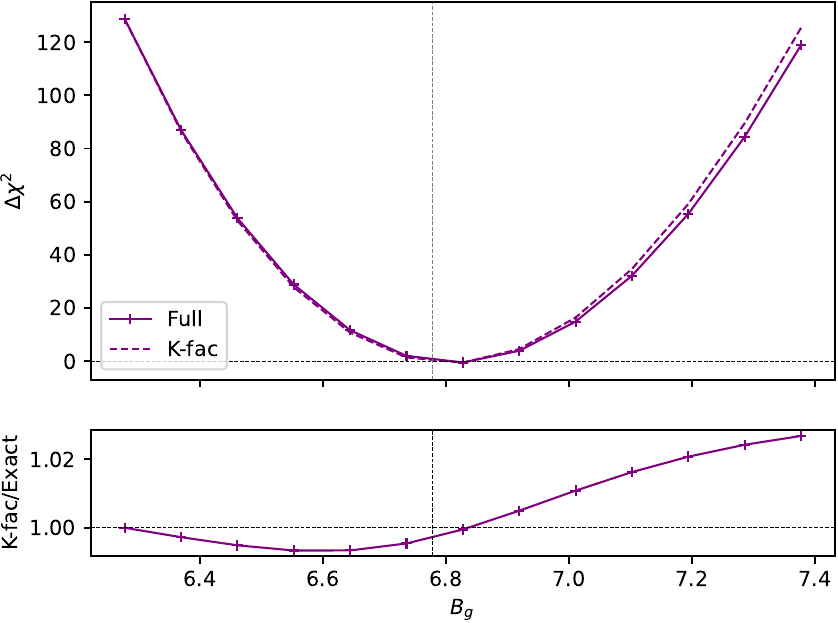} 
    \includegraphics[width=0.48\textwidth]{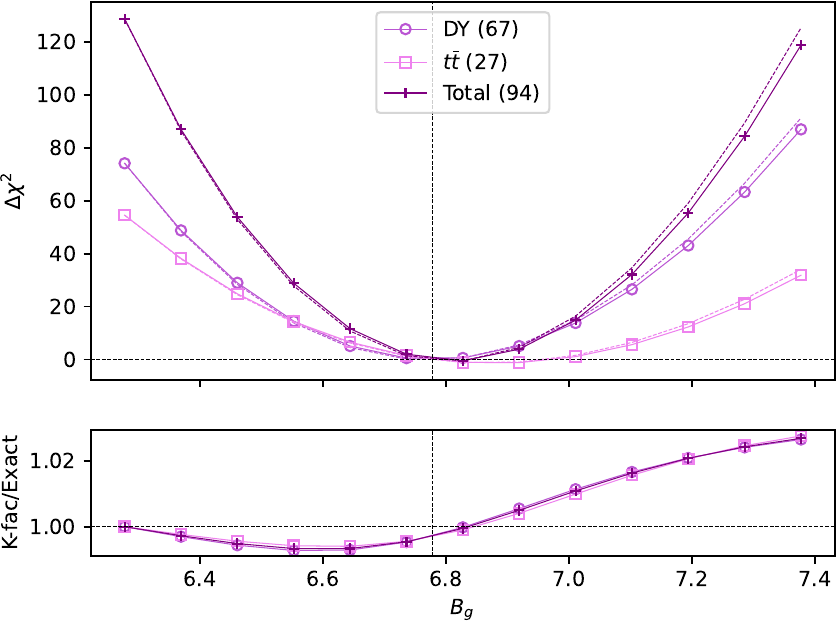} \\
    \includegraphics[width=0.48\textwidth]{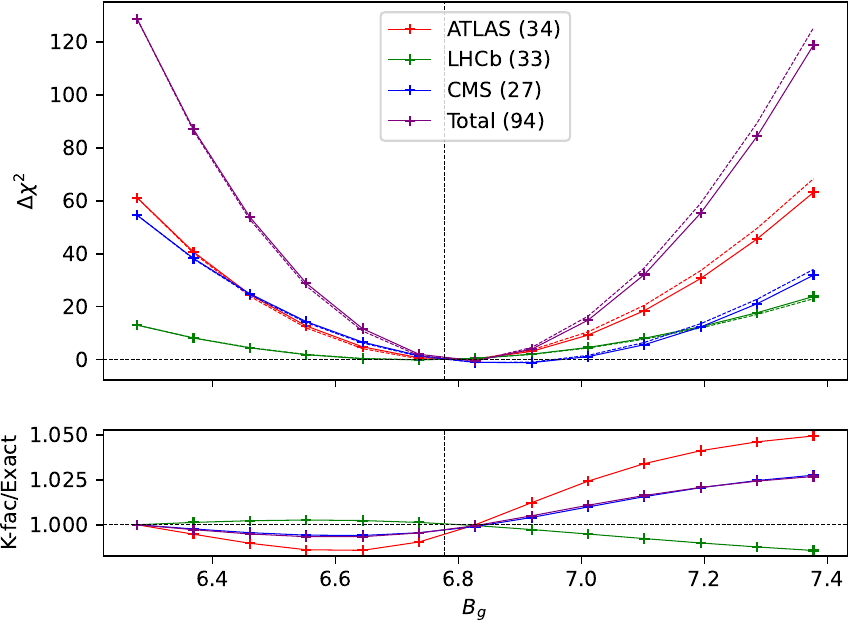} 
    \includegraphics[width=0.48\textwidth]{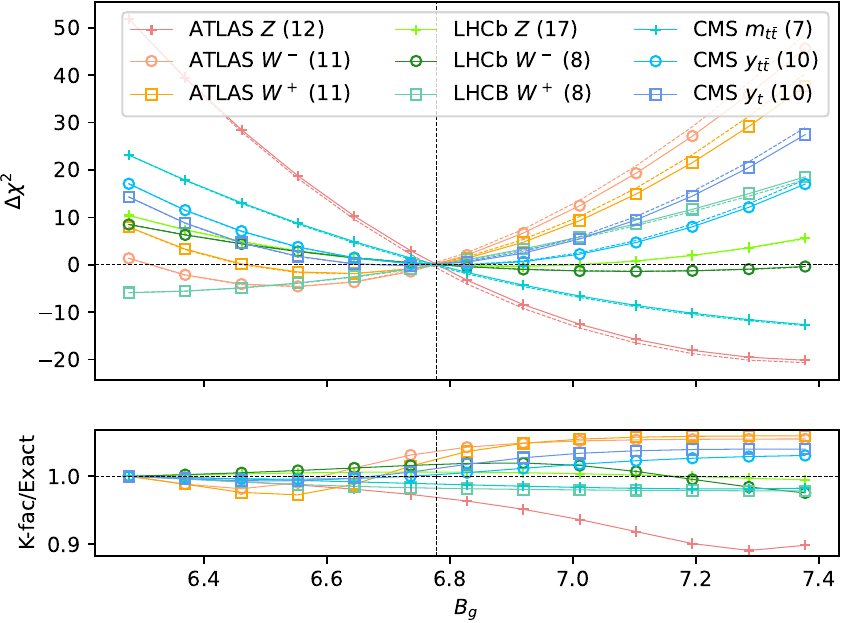} 
    \caption{Comparison of $\Delta \chi^2$ obtained using the exact NNLO calculations (solid line) versus $K$-factor rescaling (dashed line)
    for one-dimensional variations of the CJ22 gluon distribution parameter $B_g$.}
    \label{fig:scans1}
\end{figure*}

\begin{figure*}
    \centering
    \includegraphics[width=0.48\textwidth]{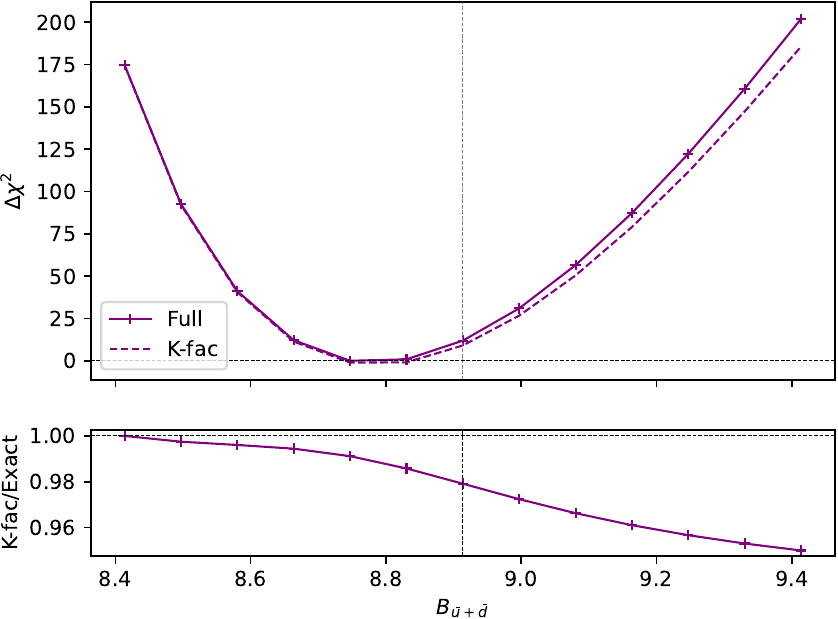} 
    \includegraphics[width=0.48\textwidth]{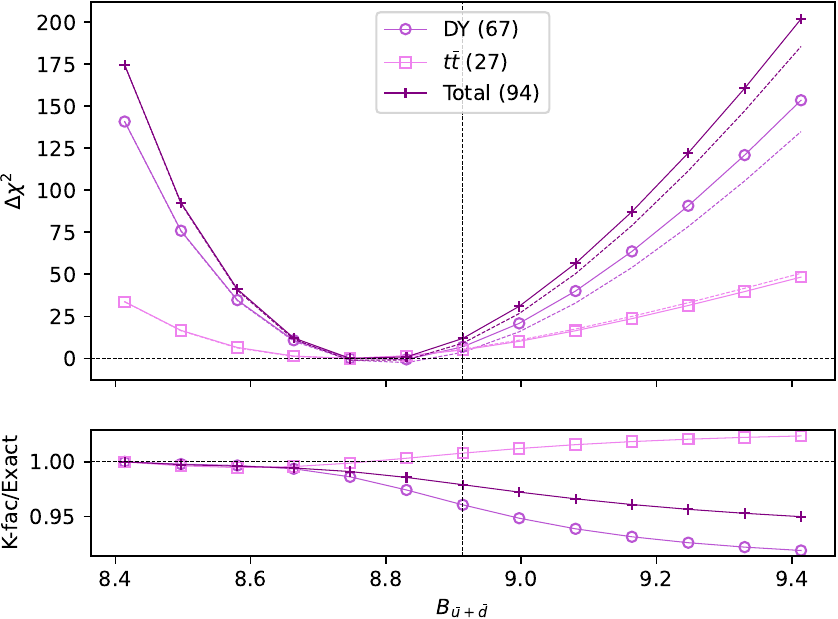} \\
    \includegraphics[width=0.48\textwidth]{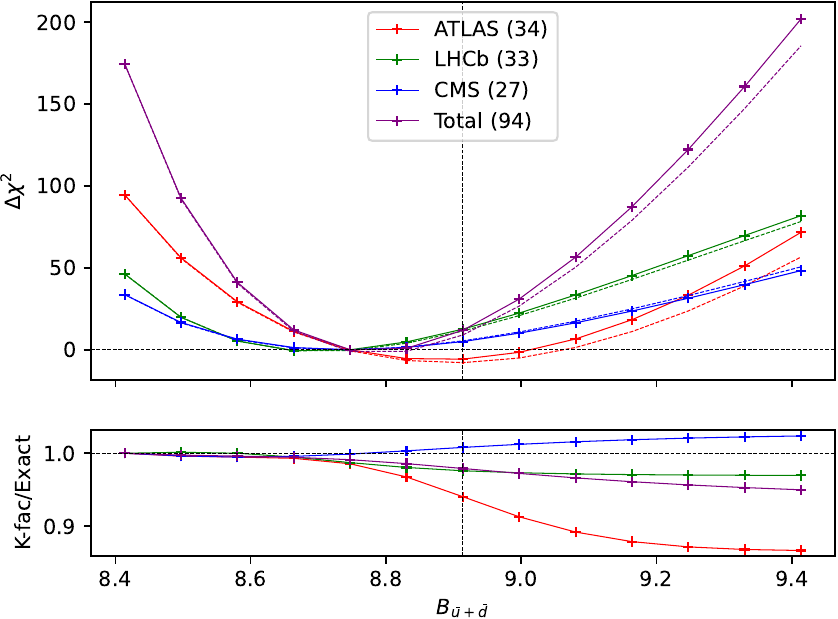} 
    \includegraphics[width=0.48\textwidth]{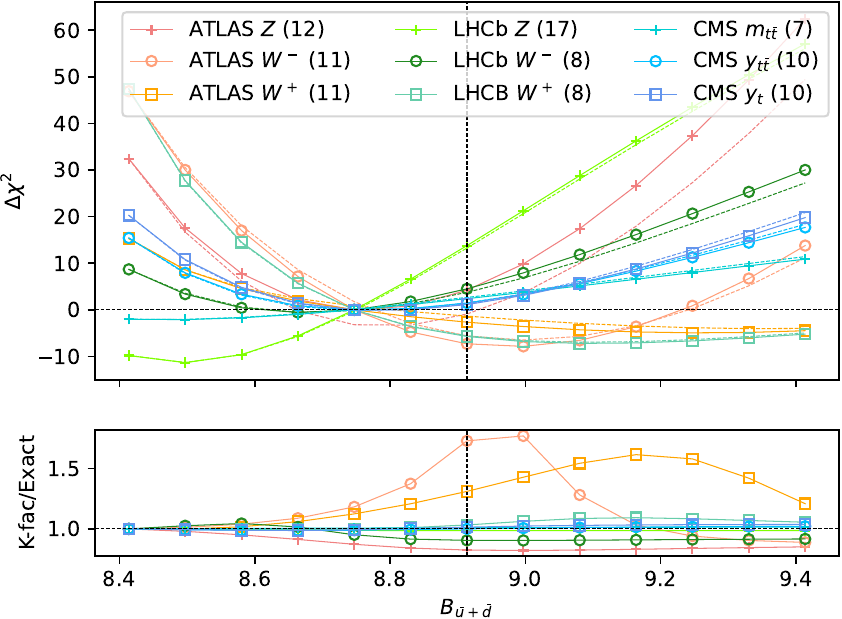} 
    \caption{Comparison of $\Delta \chi^2$ obtained using the exact NNLO calculations (solid line) versus $K$-factor rescaling (dashed line)
    for one-dimensional variations of the CJ22 $\bar{u} + \bar{d}$ distribution parameter $B_{\bar{u} + \bar{d}}$.}
    \label{fig:scans2}
\end{figure*}

The CJ22 global PDF analysis is performed at NLO QCD accuracy and includes data from a range of fixed-target experiments and from hadron colliders with lower-energy collisions.
Crucially, CJ22 does not incorporate any data from the LHC.
It employs a flexible functional form for the parametrization of each parton distribution at the input scale $Q_0$ with the gluon distribution and the sum of the $\bar{u}$ and $\bar{d}$ distributions being parametrized as:
\begin{equation}
  f(x) = A_f x^{B_f} (1-x)^{C_f} \left(1 + D_f \sqrt{x} + E_f x\right).
\end{equation}
Here, $f \in \{g, \bar{u}+\bar{d}\}$, $x$ is the momentum fraction, and $A_f$, $B_f$, $C_f$, $D_f$, and $E_f$ are free parameters that are determined by fitting.
We evolve the CJ22 PDFs in $Q^2$ using the DGLAP evolution, as implemented in \textsc{Hoppet}~\cite{Salam:2008qg}, at NNLO QCD accuracy, where the Runge--Kutta solution for the running of $\alpha_S$ is employed.\footnote{This deviates from the $\alpha_S$ and DGLAP evolution in the CJ22 analysis, but we do not worry about that here because CJ22 simply fulfils the role of a test PDF.}

In \Cref{fig:scans1,fig:scans2} we show the one-dimensional $\chi^2$ scans of the $B$ parameter, controlling the low-$x$ behaviour at the input scale, for the gluon distribution, $B_g$, and for the sum of $\bar{u}$ and $\bar{d}$, $B_{\bar{u}+\bar{d}}$, keeping all other parameters fixed.
Each figure shows four plots with $\Delta \chi^2$ as a function of the parameter value in total (top--left) or broken down by process (top--right), by experiment (bottom--left) and by measurement/spectrum (bottom--right).
In turn, each plot compares $\Delta \chi^2$ obtained using the exact predictions (solid lines) and those within the $K$-factor--based approach (dashed lines), with the ratio of the two displayed in the lower panels.
The $K$-factors are evaluated in \mbox{$B_g$ = 6.28} in \Cref{fig:scans1} and in \mbox{$B_{\bar{u} + \bar{d}}$ = 8.41} in
\Cref{fig:scans2}, with all the other values of input parameters unmodified w.r.t.\ the CJ22 fit.

The PDF input parameters are varied in a range estimated to reach a total $\Delta \chi^2$ of about 100 units, which corresponds to about \mbox{$\Delta \chi^2 \sim 1$} per data point.
This range plays a crucial role here, since the further away we get from the point where the $K$-factor was evaluated, the more potential there is for the $K$-factor--based prediction to deviate from the exact NNLO prediction.
On the one hand, a difference of \mbox{$\Delta \chi^2 \sim 1 $} per data point is much too large as compared to $\Delta \chi^2$ used to estimate Hessian-style PDF uncertainties within a single PDF fit.
On the other hand, a difference of such magnitude is not unimaginable, and possibly even too conservative if, e.g., $K$-factors calculated using a PDF from one group are reused in a fit of another group a decade later.
Here we would like to consider both perspectives, while adhering to a single test PDF. Thus we will pay attention to the local behaviour, at the scales of roughly the difference between the CJ22 fit and the minimum of the total $\chi^2$ profile, as well as to the behaviour in the whole range.

First, we observe that the CJ22 fit, indicated by the vertical dashed line, is relatively close the minimum of the profile of the total $\chi^2$ where our data would be described optimally.
Next, we note that the ratio $K$-factor/Exact, shown in the lower panel of each subplot, equals one for the lowest value in the considered ranges of the $B_g$ and $B_{\bar{u} + \bar{d}}$ parameters. 
This is expected, since these are the locations where the $K$-factors have been evaluated.
At a first glance, the choice of those locations may appear somewhat extreme.
However, evaluating the $K$-factor elsewhere in the available range would merely shift the ratio curve up or down. 
Depending on the shape, this would not reduce the difference between the two predictions by more than a factor of two (e.g. from 4\% to 2\% in the top--left ratio panel of \Cref{fig:scans2}). 
Also note that the positions where the $K$-factors are evaluated differ for the two parameter scans in \Cref{fig:scans1,fig:scans2}. 
This typically would not occur in a single PDF fit, but here allows us to probe two different scenarios.

The ratio panel on top--left plots of both figures shows that the predictions based on the full calculation and on the $K$-factors align relatively well.
Deviations are at most 1\% in the near vicinity of the minima of the $\Delta\chi^2$ profiles.
Depending on the detailed shape of these deviations, we expect the best fit locations obtained in fits relying on the full and the $K$-factor--based calculations to differ at most at the percent level, owing to the small magnitude of the correction and the quadratic nature of the $\chi^2$ profile around the minimum.
Zooming out to the full range, we find larger deviations of up to $\sim$5\%. 
Depending on whether such deviation is close to the position of the minimum or of the $\Delta\chi^2$ tolerance, this could result in shifts of similar magnitude either in the central prediction or in the uncertainty.

When inspecting the breakdowns, we observe that the DY datasets contribute more than the $t\bar{t}$ data sets. 
This may simply be due to the smaller number of $t\bar{t}$ data points than of DY.
What could also matter is the typical size of the $K$-factor compared to the precision of the measurement. 
The $K$-factors in $t\bar{t}$ are indeed larger, but so are the data uncertainties.
Furthermore, within the DY class the \ATLAS{} measurement, where the uncertainties tend to be slightly smaller than in \LHCb{}, seems to be more sensitive to the gluon parameter variations. 
The $B_{\bar{u}+\bar{d}}$ parameter is sensitive to both, with LHCb preferring the lower values and ATLAS the higher values.

\begin{figure}
    \centering
    \includegraphics[height=0.31\textheight]{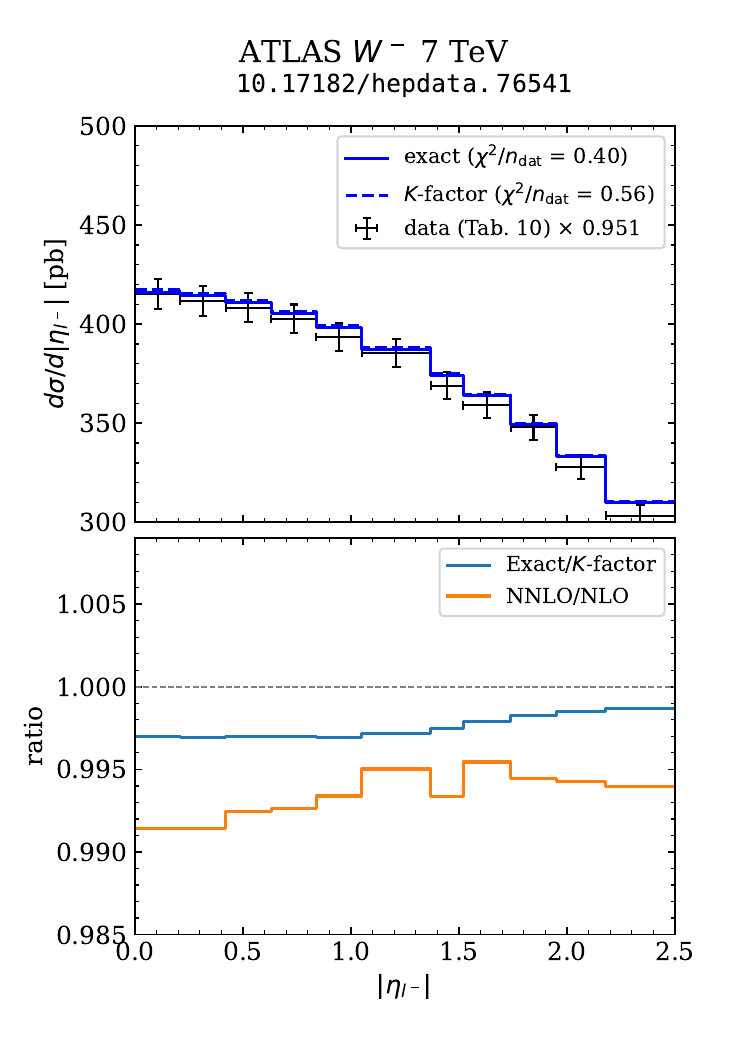} 
    \caption{Comparison of theoretical predictions obtained using the exact NNLO calculations versus $K$-factor.}
    \label{fig:ScanPredictionComparison}
\end{figure}

Digging in even deeper, we note that the different measurements have markedly different preferences for the optimal PDF for both parameters, see for example the ATLAS $W^-$ and LHCb $Z$ $\Delta \chi^2$ profiles of the $B_{\bar{u}+\bar{d}}$. 
Moreover, the data sets contributing most to the $\Delta \chi^2$ are ATLAS $Z$ and ATLAS $W^-$ for the gluon parameter, but for the \mbox{$\bar{u}+\bar{d}$} parameter the LHCb $Z$ and LHCb $W^+$ measurements play the major role. 
For those parameters and measurements the exact and $K$-factor predictions can differ by tens or even few tens of percent,
which is significant.
In order to better understand the origin of such a large difference, we compare the two theoretical predictions for one spectrum, ATLAS $W^-$, and for one variation of the parameter, $B_{\bar{u}+\bar{d}}=9.41$, which corresponds to the upper edge of the parameter scan in \Cref{fig:scans2}.
We find, see \Cref{fig:ScanPredictionComparison}, that the difference between the exact and $K$-factor predictions (plotted in blue) for this choice of PDF is of the same order of magnitude as the actual NNLO $K$-factor (plotted in orange). 
Thus the magnitude of the $K$-factor PDF dependence in the selected parameter variation range is not negligible as compared to its typical size.

Overall, we find that the goal precision of 1\%, in the parameter space, for estimating the position of the minimum can be achieved if particular care is taken with respect to the position where the $K$-factor is evaluated. 
This should already be the case if its final value is estimated close to the position of the new best fit.
In that case also the uncertainties, the goal precision of which can be relaxed, can likely be estimated reliably. 
Further away from this ideal location this is no longer the case. 
There we found instances where the variation of the $K$-factor due to its PDF dependence is of similar size as the $K$-factor itself and thus cannot be neglected.
When considering all measurements at once, the deviations of the data description between the exact and the $K$-factor predictions we observed were not very large, but this is not an implication of the fact that the $K$-factor and exact calculations were in satisfactory agreement in all measurements, but rather that their deviations are ``washed out'' by the difference in relative pulls across different measurements.
Thus, the reliability of $K$-factors should not be generally assumed in NNLO fits. 
In fits at even higher formal accuracy, like N3LO, the use of $K$-factors should be done with prudence.

\subsection{NLO EW corrections in interpolation grids\label{sec:interpolation-grids-at-nnlo-qcd-nlo-ew}}

\begin{figure*}[t]
  \centering
  \includegraphics[height=0.18\textheight]{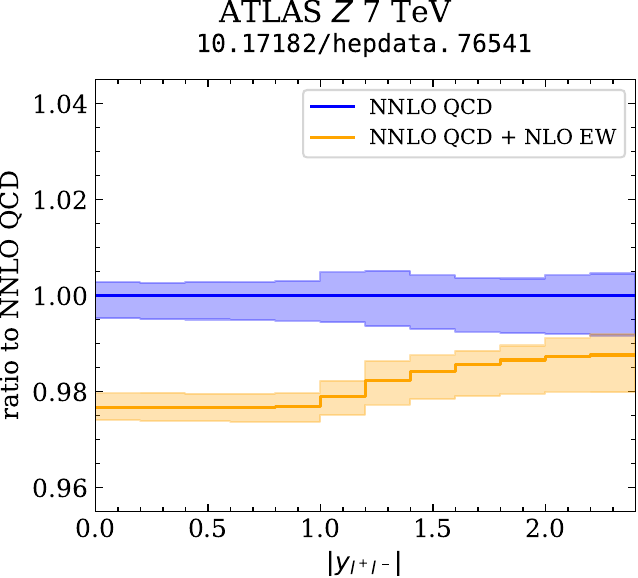}\hfill
  \includegraphics[height=0.18\textheight]{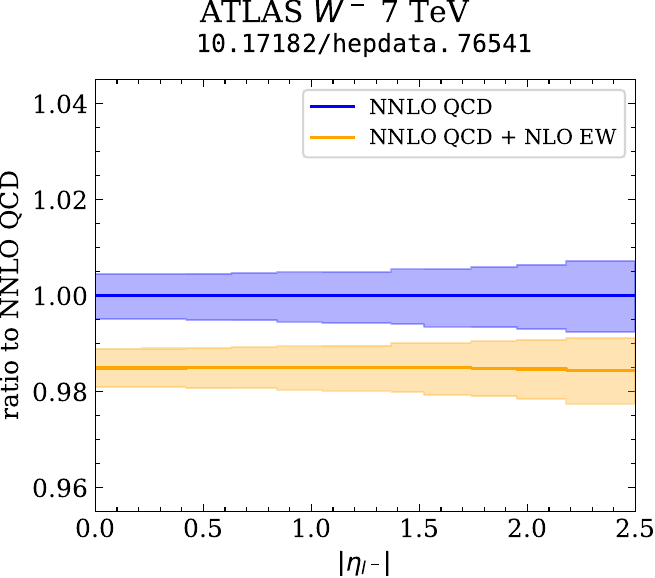}\hfill
  \includegraphics[height=0.18\textheight]{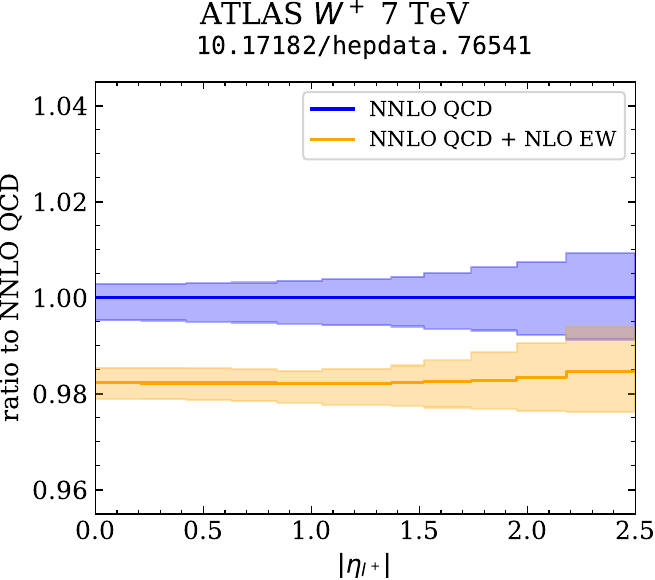}\\[2mm]
  \includegraphics[height=0.18\textheight]{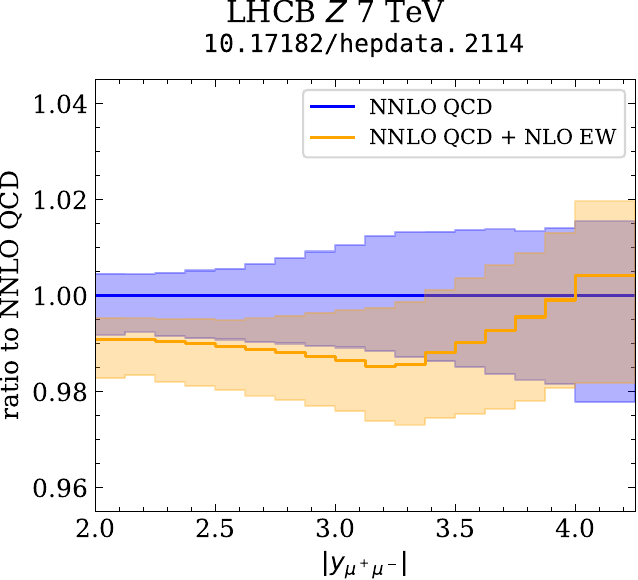}\hfill 
  \includegraphics[height=0.18\textheight]{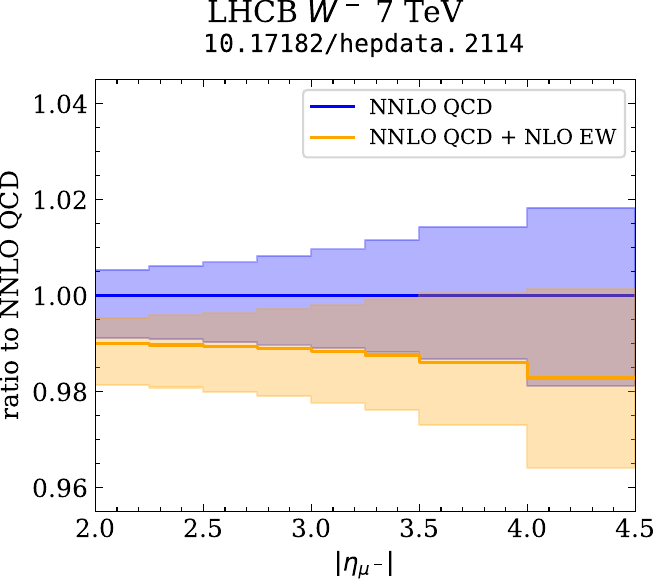}\hfill
  \includegraphics[height=0.18\textheight]{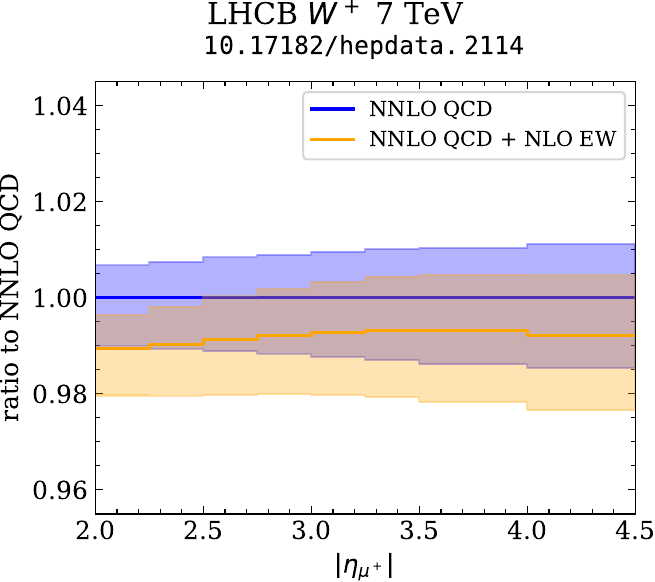}
  \caption{Electroweak corrections for the ATLAS (upper row) and LHCb (lower row) measurements of $Z$ and $W^\pm$ production~\cite{ATLAS:2016nqi,LHCb:2015okr}. The NNLO QCD+NLO EW predictions (in orange) are normalized by the NNLO QCD prediction (in blue). Both predictions feature 7-pt QCD scale variation bands.}
  \label{fig:electroweak}
\end{figure*}

\PineAPPL{} and \Matrix{} both support the simultaneous expansion in the strong and electroweak couplings and as such enable the inclusion of higher-order EW effects.
To demonstrate this feature, we show the impact of the EW corrections on our predictions for the DY measurements discussed before.
In \Cref{fig:electroweak} we depict the size of the NLO EW corrections (orange) relative to the NNLO QCD predictions (blue) with their respective 7-pt QCD scale uncertainties.\footnote{The predictions have been computed by using an additive approach for the combination of QCD and EW corrections. Multiplicative approaches could be obtained from the same grids since each perturbative order is stored separately therein, but some care should be taken~\cite{Grazzini:2019jkl}.} 
We observe that, in particular in the central-rapidity regions (see predictions for the ATLAS setup in the upper row), the EW corrections can appreciably exceed the NNLO QCD uncertainties.
We find that the EW corrections are negative and at the level of few percent, varying by at most $1\%$ over the considered ranges.
Such effects are, however, not small compared to the respective perturbative uncertainties which are at or even below the percent level at NNLO QCD accuracy.
In the forward-rapidity regions (see predictions for the LHCb setup in the lower row), shape effects tend to be more pronounced, but are usually covered by the larger QCD scale uncertainties.
An even more distinct phase space dependence is expected if energy-dependent distributions, whose tails are dominated by EW Sudakov logarithms, are considered.

\begin{figure*}[h]
  \centering
  \includegraphics[height=0.31\textheight]{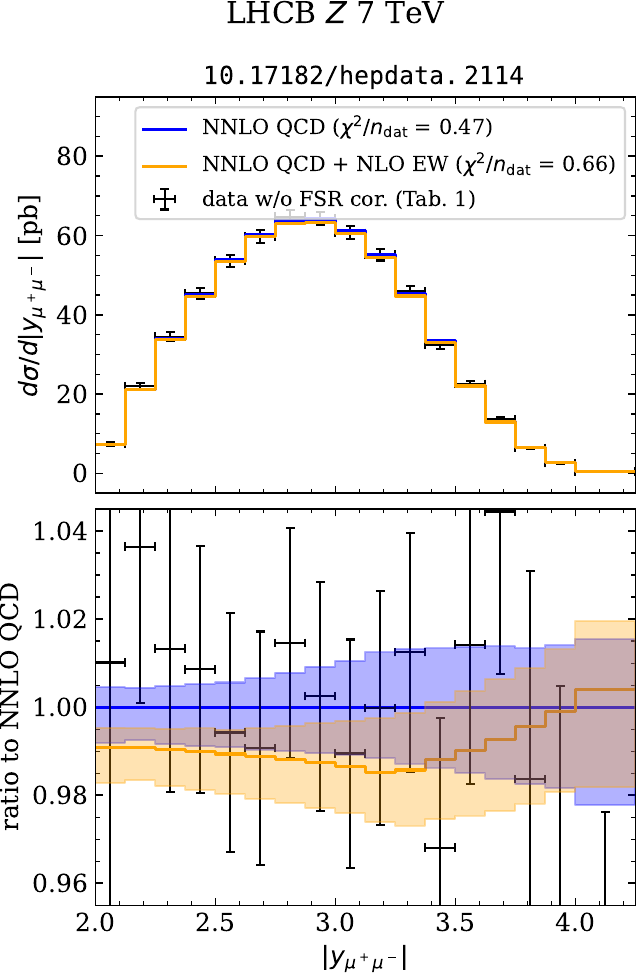}\hfill 
  \includegraphics[height=0.31\textheight]{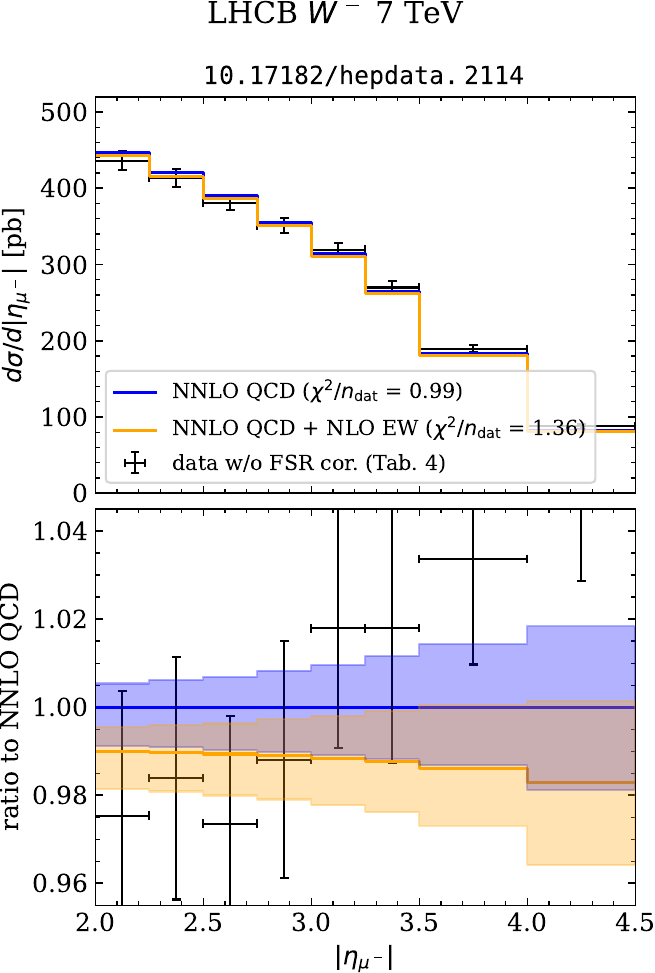}\hfill
  \includegraphics[height=0.31\textheight]{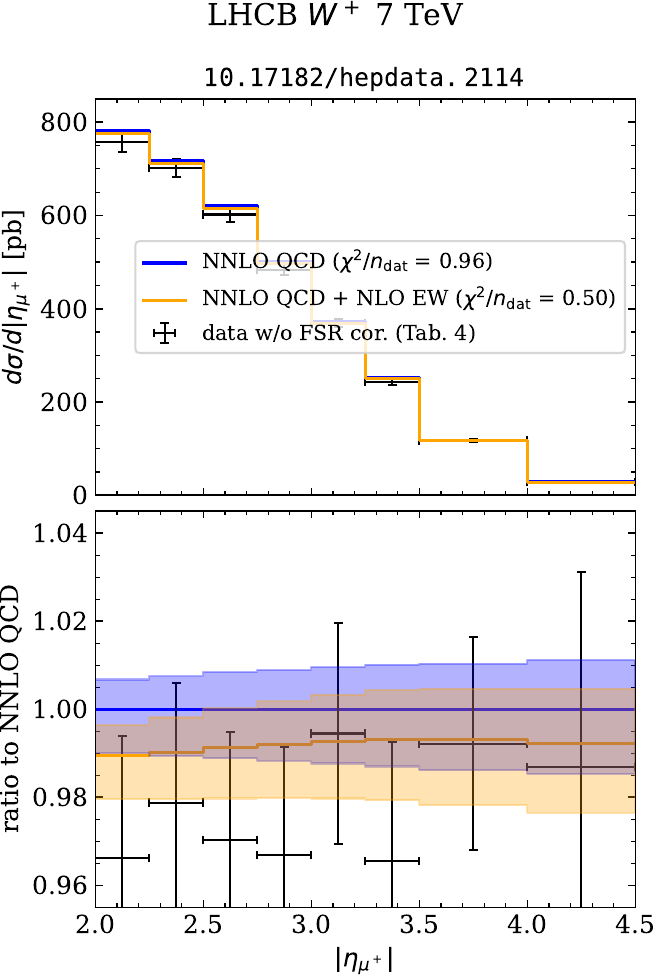}
  \caption{NNLO QCD+NLO EW predictions (orange) compared to data from the LHCb measurement~\LHCbCitation{} divided by the FSR correction factor. For reference we also show the NNLO QCD predictions (blue) since previously they were compared to data that included FSR correction factors and was normalized such as to minimize the $\chi^2/n_{\text{dat}}$. 7-pt scale uncertainties are also shown.}
  \label{fig:electroweak2}
\end{figure*}

In \Cref{fig:electroweak2} we compare the NNLO QCD+NLO EW predictions to the LHCb data~\LHCbCitation{} where the QED FSR correction factors\footnote{These correction factors can be found in the third column of Table 1, for $Z$ boson production, and in third and fifth columns of Table 4, for $W^+$ and $W^-$ boson production, respectively.}, originally applied in the analysis, have been divided out. 
While this measurement is not yet precise enough to fully benefit from them, EW corrections are very likely going to matter in the upcoming measurements of $Z$ and $W^\pm$ boson production at the LHC.

In the context of EW corrections, some remarks about the treatment of the CKM matrix are in place.
In general, it is set to unity in \Matrix{}, since the expected effects of a non-trivial CKM matrix may usually be assumed to be negligible compared to other uncertainties.
The only exception at present is charged-current DY production, given the high accuracy on the experimental side. Here, \Matrix{} allows a general CKM matrix to be used, with the common limitation of most Monte Carlo generators that, if NLO EW corrections are included, this general CKM matrix cannot be applied throughout the entire calculation:
The reason is that, to date, the amplitude providers \Matrix{} relies on do not support a proper renormalisation of the non-trivial CKM matrix, as it would be required to compute the corresponding EW loop amplitudes exactly.
The standard strategy in \Matrix{} is thus to approximate either the complete or only the virtual EW corrections with a trivial CKM matrix and to keep the exact dependence in the rest of the calculation.
The impact of this approach is usually of the order of per mille at the level of predictions for LHC physics, and thus a sufficiently good approximation. 
However, for PDF determinations the impact of such approximation might not be negligible, since the CKM matrix elements distribute the cross sections among different PDFs.
We have thus added the feature to restore at least the Born-like CKM factors, based on the respective incoming quark--anti-quark pairs, in a reweighting
approach, while performing the actual loop calculation with a trivial CKM matrix.\footnote{This feature can now be accessed in \Matrix{} by selecting in the input file \texttt{parameter.dat} the option \texttt{approx\_ckm\_EW = 0}.}
The grids produced in this publication and published on \texttt{PloughShare} rely on this approach.

\subsection{PDF variations and uncertainties\label{sec:PDFunc}}

\begin{figure*}
  \centering
  \includegraphics[width=\textwidth]{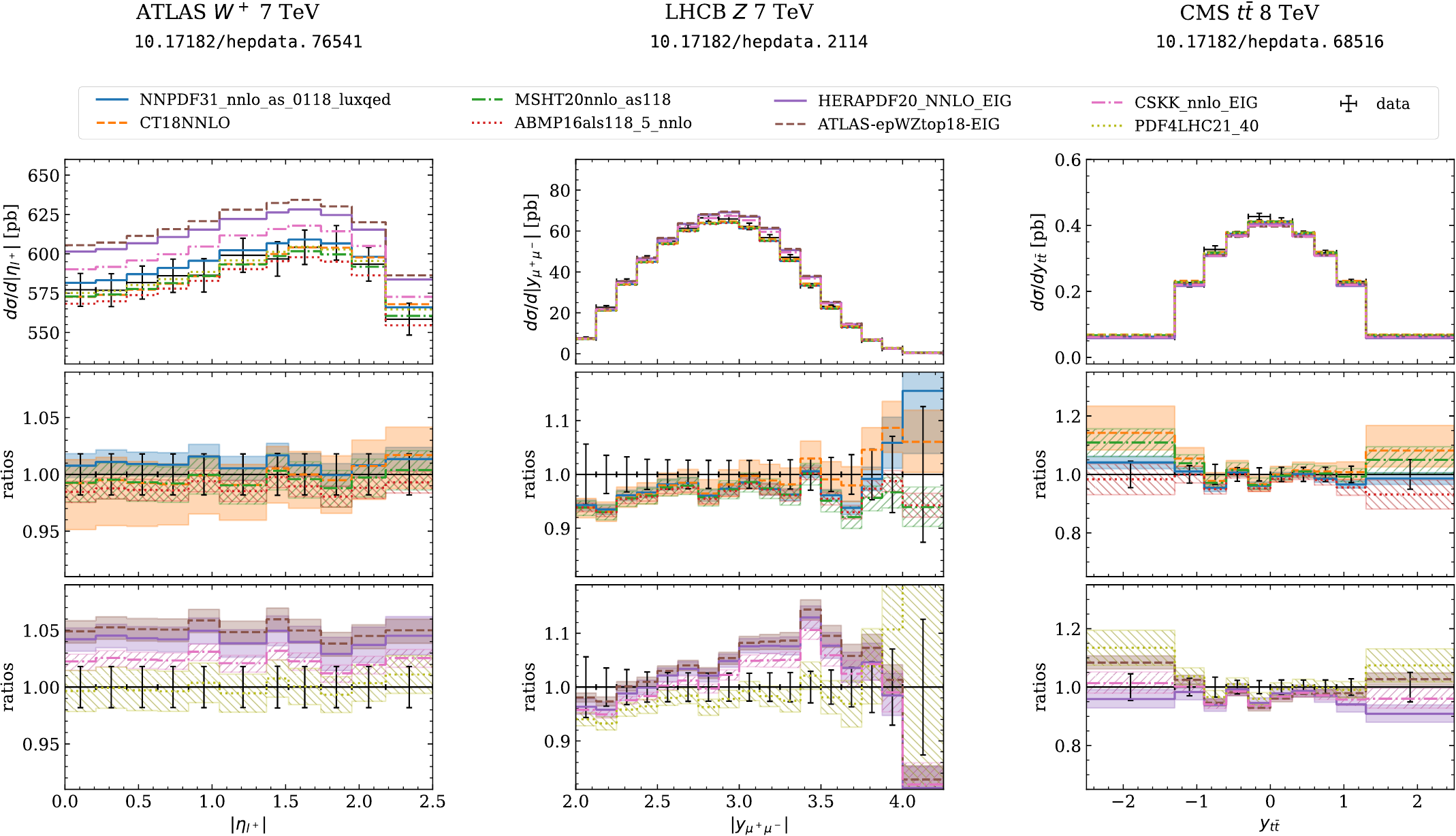}
  \caption{Example predictions for a subset of the previously discussed measurements from ATLAS, LHCB and CMS for a selection of
    recent PDF sets including PDF uncertainties.}
  \label{fig:PDF}
\end{figure*}

With the interpolation grids calculated in the previous sections we now have the possibility to easily vary the PDF sets in theoretical predictions a posteriori.
We can use this feature to thoroughly assess the PDF uncertainties of the theoretical predictions for these data sets.
This can be done by varying PDF members and combining them into an uncertainty band according to the prescription appropriate for a given PDF set and by employing PDF sets from different groups in order to uncover possible tensions.

In \Cref{fig:PDF} we show comparisons of our predictions for 8 different PDF sets (NNPDF31, CT18, MSHT20, ABMP, HERAPDF, ATLAS, CSKK and PDF4LHC) in the upper panel.
In the lower panels we show the ratios w.r.t.\ data of those predictions including PDF uncertainties.
For the sake of readability, we split the ratios into two panels.

\section{Conclusions}
\label{sec:conclusions}
Near the high-luminosity phase of the LHC an unprecedented level of precision and accuracy is required for basically all SM processes. 
To achieve this, besides pushing forward the frontier of scattering processes that can be described at NNLO QCD and beyond, a fundamental need is to improve our knowledge of PDFs, which in turn affect the achievable accuracy level of precision calculations where they enter as non-perturbative input.
Moreover, not only do such calculations need to improve in precision, they also have to provide reliable estimates of the uncertainties on their predictions. 
While providing scale variation uncertainties as (a lower bound of) residual perturbative uncertainties at a given order has become standard, it is typically not as easy to investigate PDF uncertainties for involved higher-order calculations. 
The latter is often due to code structures that are not well suited for evaluating results for a larger number of PDF sets (or members of error PDF sets) without re-doing the whole calculation several times.
This holds true also for the \Matrix{} framework, which is a public tool that is able to calculate NNLO QCD and NLO EW corrections for a wide range of processes, but, as a standalone code, cannot calculate PDF uncertainties in a practical manner.

While this functionality could in principle be added directly in some way or the other, this paper follows a different path that provides solutions for both of the aforementioned essential tasks: we have established an interface between \Matrix{} and \PineAPPL{}, which is a tool built to generate and deal with so-called interpolation grids in a highly efficient way.
Such interpolation grids provide a format to store the outcome of calculations for differential cross sections up to in principle arbitrary orders in the strong and electroweak couplings independent of an explicit PDF set.
The convolution with any PDF set can then be performed a-posteriori and almost instantaneously\footnote{We set up a speed test on a compute node of the PALMAII cluster at the University of M\"unster in which we convolve the 9 NNLO grids published here, with a total of 94 bins, with 9 different PDF sets using the PineAPPL command line interface. 
All measurements were performed on an Intel(R) Xeon(R) Gold 6140 CPU @ 2.30 GHz and repeated 20 times, with the reported numbers corresponding to the averages. The runtime measured by the CLI includes the time required to read the grids. To estimate the marginal convolution cost separately from this overhead, we measure the runtime once using a single PDF set and once using nine PDF sets, and extract the additional cost per convolution from the difference. Averaging over bins yields typical convolution times of about 40–100 ms per bin.}, 
which paves the way for the evaluation of all kinds of PDF uncertainties, as well as for easily updating results from once performed precision calculations with newly available PDF sets.
Moreover, the repeated convolution of full higher-order calculation results with iteratively adapted PDF sets, which is a key ingredient of PDF determination procedures, becomes straightforwardly accessible. 
In particular the latter is completely impractical without the interpolation grid approach, but with the new interface between \Matrix{} and \PineAPPL{}, which we named \MatrixHawaii{}, interpolation grids become a standard output format for \Matrix{} and can be immediately used for all these operations.

We have discussed the required steps to preserve all important features of \Matrix{} results in the \PineAPPL{} interpolation grids delivered by \MatrixHawaii{}; worth noting in particular is how the extrapolation procedure used to get rid of the slicing parameter
dependence of our approach is promoted to the grids.
Moreover, we have presented a dedicated validation for a series of processes and fiducial setups, chosen according to their relevance for PDF determination, to show that interpolation errors are under control. 
These errors are an inevitable drawback of the interpolation grid approach, but we have illustrated that they are far below the numerical precision we can achieve within reasonable runtimes at NNLO QCD, i.e.\ well below the per mille level for distributions, and yet much smaller than other relevant sources of uncertainties like residual perturbative uncertainties or those from PDFs.

Eventually, we have presented some practical applications, including studies of PDF uncertainties and on the inclusion of NLO EW corrections on top of
NNLO QCD through the interpolation grids.
Those are performed for the aforementioned sample processes and setups used in our validations.
More precisely, these sample processes are Drell--Yan production at \SI{7}{\tera\electronvolt} in the setups applied by ATLAS~\ATLASCitation{} and LHCb~\LHCbCitation{}, respectively, as well as top-quark pair production at \SI{8}{\tera\electronvolt} used by CMS~\CMSCitation{}. 
The NNLO-accurate grids in the \PineAPPL{} format that we have generated with \MatrixHawaii{}, which the results presented in this publication are based on, are made public on {\tt PloughShare}.

We have further used these sample sets to study the impact on PDF determinations of following the commonly used practice of taking NNLO QCD predictions into account only through $K$-factors instead of applying interpolation grids with exact NNLO QCD information encoded.
In particular, the $K$-factor approach neglects the dependence of NNLO corrections on the respective partonic channels and, moreover, introduces a dependence on how exactly those $K$-factors are evaluated, in particular the choice of the PDF set used.
We did so in a simplified approach --- performing full PDF fits in both approaches and comparing the outcome would have been beyond the scope of this publication.
Instead we studied how the data quality description changes when switching between exact and $K$-factor--based predictions as  we depart from a minimum of a single test PDF fit. 
We find that it may certainly not be taken for granted that the approximate $K$-factor approach comes without a bias on the PDF sets determined, although it seems to perform quite well in several situations.
Nevertheless, since \MatrixHawaii{} is released with this publication, its feature to generate NNLO-accurate \PineAPPL{} interpolation grids for all processes available in the \Matrix{} framework --- Drell--Yan, Higgs, diboson, triphoton and top-pair production by now --- should make a $K$-factor approach in PDF determination obsolete for these process classes.

\bmhead{Acknowledgements}
We would like to thank Massimiliano Grazzini and Aleksander Kusina for insightful discussions and comments on the manuscript.
This research was supported by the Munich Institute for Astro-, Particle and BioPhysics (MIAPbP) which is funded by the Deutsche Forschungsgemeinschaft (DFG, German Research Foundation) under Germany's Excellence Strategy - EXC-2094 - 390783311.
The work of S.D.\ has been funded by the European Union (ERC, MultiScaleAmp, Grant Agreement No.\ 101078449).
Views and opinions expressed are however those of the author(s) only and do not necessarily reflect those of the European Union or the European Research Council Executive Agency.
Neither the European Union nor the granting authority can be held responsible for them.
C.S.\ acknowledges financial support by the German Federal Ministry for Education and Research (BMBF) under contract no.\ 05H21WWCAA and the German Research Foundation (DFG) under reference number DE 623/8-1.
The work of T.J.\ was partly funded by the SFB 1225 ``Isoquant'', project-id 273811115. 
Computational resources for this research were in part provided by SMU’s O’Donnell Data Science and Research Computing Institute.

\appendix

\section{Manual of \MatrixHawaii{}}\label{app:manual}

\Matrix{} is a public computational framework, and its latest release can be downloaded from its website: {\tt https://matrix.hepforge.org/}. 
While the future releases of \Matrix{} will include the interface with \PineAPPL{}, this is not supported by its current public version ({\tt v2.1.0}).
We thus provide a beta version of the interface between \Matrix{} and \PineAPPL{} ({\tt v2.2.0.beta}), i.e.\ \MatrixHawaii{}, on the website with the submission of this paper.\footnote{Together with the \PineAPPL{} interface, version {\tt v2.2.0.beta} includes the fix of a bug in the random number management, which had affected certain (pseudo-)rapidity distributions, in particular in low multiplicity processes. Besides, an improved treatment of the CKM matrix in NLO EW corrections is now available through the setting \texttt{approx\_ckm\_EW = 0}, as discussed in detail in \cref{sec:interpolation-grids-at-nnlo-qcd-nlo-ew}.}

This Appendix provides a manual for the usage of such interface, and is thought as a supplement to the \Matrix{} user manual\footnote{The \Matrix{} user manual can be downloaded from {\tt https://matrix.hepforge.org/manual.html}.}, to which we refer the reader for any additional query regarding the code.

\subsection{Installation and compilation}

The configuration of \Matrix{} follows the standard procedure described in the \Matrix{} user manual, with the addition of the link to the \PineAPPL{} installation.
In order to generate interpolation grids, \Matrix{} requires the installation of the C-language interface (CAPI) of \PineAPPL{}.
If a local installation is available, the user can specify the path to it in the file {\tt MATRIX\_configuration}.
If no path is provided, the script automatically downloads a precompiled version of the \PineAPPL{} CAPI and links it to \Matrix{}.

The compilation of a process also follows the standard procedure, and the connection to \PineAPPL{} is established via an additional argument.
Thus, the user only interested in the original implementation of \Matrix{}, by running the \Matrix{} script in the standard way will not install any related software and will not notice any difference in the usage of \Matrix{}.
If the user is instead interested in the \PineAPPL{} interface, they need to compile the process with the additional flag {\tt ---hawaii}:\footnote{In the unlikely case that switching between compilations with and without the flag {\tt ---hawaii} is required, the code must be recompiled from scratch.}
{\tt
\begin{lstlisting}[language=bash]
 $ ./matrix ${process_id} --hawaii
\end{lstlisting}
}

\subsection{Running a process}

After compilation, a process can be run as described in the \Matrix{} user manual from the \Matrix{} process folder (default: {\tt run/\$\{process\_id\}\_MATRIX}).

The only difference appearing in the settings of the run is an additional switch in the file {\tt parameter.dat} that allows the user to turn on or off the generation of \PineAPPL{} interpolation grids:
{\tt
\begin{lstlisting}[basicstyle=\scriptsize,frame=single]
switch_PineAPPL  =  1          
\end{lstlisting}
}

After the end of the main run, the summary routine of the \Matrix{} script automatically performs the summary also of the \PineAPPL{} grids for the total cross section and each single- or double-differential distribution.\footnote{Note that the \PineAPPL{} grids require a considerable amount of memory, with memory usage scaling proportionally to the number of non-trivial bins.}
The resulting grids are copied into the result folder:
{\tt
\begin{lstlisting}[language=bash, basicstyle=\scriptsize]
  run/{process_id}_MATRIX/result/
   run_{run_id}/{order_id}-run/
   PineAPPL_grids/
\end{lstlisting}
}

The interpolation grids generated in this way further store additional information in the form of metadata.
This metadata can be read, for instance, by using the {\tt read} command of the \PineAPPL{} command line interface (CLI):
{\tt
\begin{lstlisting}[language=bash]
  pineappl read --show 
   {grid_name}.lz4
\end{lstlisting}
}

In the following, we provide a list of the metadata automatically filled by \Matrix{}.

\begin{itemize}
\item {\tt citations}: a copy of the {\tt CITATION.bib} file generated by \Matrix{}, where all the publications relevant for the run that produced the grid are listed. Please cite these papers if you use the results from \Matrix{}, to acknowledge the work that went into obtaining them.
\item {\tt parameter.dat}: a copy of the input file {\tt parameter.dat} of the run;
\item {\tt model.dat}: a copy of the input file {\tt model.dat} of the run;
\item {\tt distribution.dat}: a copy of the input file {\tt distribution.dat} of the run;
\item {\tt dddistribution.dat}: a copy of the input file {\tt dddistribution.dat} of the run (if it exists);
\item {\tt runtime.dat}: a copy of the {\tt runtime.dat} file, required in case it is desired to exactly reproduce the result of the run, i.e.\ using identical phase space points throughout;
\item {\tt LHAPDFname}: PDF set used in the main \Matrix{} run;
\item {\tt LHAPDFsubset}: index of the PDF set's member used in the main \Matrix{} run;
\item {\tt results}: a table showing the results for each bin of the differential cross section 1) from \PineAPPL{}, which is the interpolated value from the 2) \Matrix{} result.
The next column shows the 3) relative Monte Carlo integration uncertainty in percent.
The fourth column shows the interpolation error for the central scale choice once in terms of 4) the Monte Carlo uncertainty and another time 5) in terms of per mille.
The final columns show the interpolation error for 6) smallest and 7) largest scale-varied result in per mille.
This tables serves as a quick sanity check for the user; the first two columns should be the same and the remaining columns close to zero.
The interpolation error should usually not be larger than a few per mille and is often even below one per mille.
\item {\tt x1\_label}: name of the observable;
\item {\tt x1\_label\_tex}: name of the observable (\TeX\, format);
\item {\tt x1\_unit}: unit of the observable;
\item {\tt x2\_label}: (for double-differential distributions) name of the second observable;
\item {\tt x2\_label\_tex}: (for double-differential distributions) name of the second observable (\TeX\, format);
\item {\tt x2\_unit}: (for double-differential distributions) unit of the observable;
\item {\tt y\_label}: name of the differential cross section;
\item {\tt y\_label\_tex}: name of the differential cross section (\TeX\, format);
\item {\tt y\_unit}: unit of the differential cross section.
\item {\tt specify.cuts.cxx}, {\tt specify.scales.cxx}, {\tt specify.particles.cxx}, \\{\tt specify.prepare.scales.cxx}: a copy of the corresponding user-defined files. This metadata entry is filled only if the files differ from those stored in the default folder.
\end{itemize}

\section{\texorpdfstring{\boldmath$\rcut$}{r cut}-parameter extrapolation with interpolation grids\label{app:extrapolation}}

One important procedure of $q_\mathrm{T}$-subtraction slicing methods is the extrapolation of the slicing parameter \mbox{$\rcut \to 0$}, to eliminate the residual dependence on the (typically small) value of this parameter.
\Matrix{} implements this procedure in an automated fashion, and it can be enabled and disabled easily using its configuration files.
In \cref{sec:rcut-parameter-dependence} we discuss the importance and impact of this extrapolation, and here we briefly address some features of its implementation.

The ansatz to perform this extrapolation in \mbox{$\rcut \to 0$} is to calculate the cross sections $\sigma (\rcuti)$ for specific finite values of $\rcuti$, and then to fit these results to a polynomial (possibly including logarithmic enhancements) in $\rcut$.
This polynomial can then be explicitly evaluated at \mbox{$\rcut = 0$}, thereby providing the extrapolated results.

In the following subsections we show the impact of the following properties: 1) interpolation grids with the same nodes are elements of a vector space (grids can be added and multiplied with a scalar) and 2) the convolution of interpolation grids with PDFs is a linear operation.
A consequence of 1) is that the method of least squares is applicable to interpolation grids, and therefore the \mbox{$\rcut \to 0$} extrapolation procedure generalizes to interpolation grids in such a way that, had the integration been performed without interpolation grids, the same result would be obtained.
Furthermore, property 2) implies that the extrapolation procedure is independent of PDFs, in a way that has to be properly defined.

\subsection{Fitting predictions with a constant}

Let us assume that we have a set of $N$ cross sections
\begin{equation}
\left\{ \sigma_i \equiv \sigma (\rcuti) \right\}_{i=1}^N \,,
\label{eq:rcut-cross-sections}
\end{equation}
sampling the dependence of $\sigma (\rcut)$ with varying slicing parameters $\rcut$.
We can only calculate the cross sections $\sigma_i$ with non-zero \mbox{$\rcut > 0$} far enough from zero without running into numerical cancellation problems, but would like to estimate the limit
\begin{equation}
\sigma \equiv \lim_{\rcut \to 0} \sigma (\rcut)\, \text{.}
\label{eq:extrapolation}
\end{equation}
We further assume that the dependence on $\rcut$ is polynomial.
To simplify matters, let us assume for the time being that we have chosen the $\rcuti$ small enough so that there is a plateau and actually there is no dependence on $\rcut$:
\begin{equation}
\sigma (\rcut) = \beta\, \text{.}
\label{eq:constant-fit}
\end{equation}
We will relax this assumption in the next subsection, but the general findings will be unaffected by this choice.

We now use the method of least squares to fit the parameter $\beta$.
This means minimizing the sum of the square of the residuals,
\begin{equation}
S = \sum_{i=1}^N (\beta - \sigma (\rcuti))^2\, \text{,}
\label{eq:constant-fit-unweighted-leastsquares}
\end{equation}
which requires \mbox{$\partial S/\partial \beta = 0$}.
This yields the average of the cross sections:
\begin{equation}
\beta = \frac{1}{N} \sum_{i=1}^N \sigma (\rcuti)\, \text{.}
\label{eq:constant-fit-result}
\end{equation}
Now we would like to answer the following questions:
\begin{enumerate}
\item Is the fitting of the $\rcut$-dependence of $\sigma (\rcut)$ sensitive to the choice of PDF set or not?
If it is, this would be problematic from the point of view of incorporating the results into a PDF fit.
All input to a PDF fit should be independent of PDFs, the quantities that are being fitted.
\item Is there an interpolation grid corresponding to $\sigma$ as a result of the extrapolation in \cref{eq:extrapolation}?
\end{enumerate}
For the case of a constant fit, \cref{eq:constant-fit}, it is straightforward to derive answers to these questions; but first we need the connection with interpolation grids.
Let $\{ g_i \}_{i=1}^N$ be the interpolation grids corresponding to \cref{eq:rcut-cross-sections}, such that
\begin{equation}
C (g_i) = \sigma_i\, \text{,}
\label{eq:grid-convolution}
\end{equation}
where $C (g_i)$ denotes the convolution of a grid $g_i$ with a specific PDF set, yielding the cross section $\sigma_i$.
We note that convolutions $C$ are linear functions,
\begin{equation}
C (\alpha g_1 + g_2) = \alpha C(g_1) + C(g_2)\, \text{,}
\label{eq:convolution-linearity}
\end{equation}
because we can add grids with the same nodal points and multiply them with scalars $\alpha \in \mathbb{R}$.
Going back to \cref{eq:constant-fit-result}, inserting \cref{eq:grid-convolution} into it and making use of the linearity of the convolution, \cref{eq:convolution-linearity}, we find
\begin{equation}
\beta = C \left( \frac{1}{N} \sum_{i=1}^N g_i \right) \text{.}
\end{equation}
In other words, we find an interpolation grid
\begin{equation}
\tilde{g} = \frac{1}{N} \sum_{i=1}^N g_i\, \text{,}
\label{eq:grid-fit}
\end{equation}
so that $\beta = C (\tilde{g})$\,.

The existence of $\tilde{g}$ is a consequence of the fact that the $\rcut$ fit and the extrapolation \mbox{$\rcut \to 0$} is \emph{independent} of a chosen PDF fit at the level of (unconvolved) partonic cross sections, because \cref{eq:grid-fit} itself does not make use of any PDF set.

\subsection{Fitting predictions with general polynomials}

In the previous section we restricted ourselves to fit a constant. However, the generalization to, e.g., a linear function is straightforward and lets us easily generalize the procedure to a polynomial fit and beyond.
To that end, let us minimize the sum of the square of the residuals, fitting a linear function,
\begin{equation}
S = \sum_{i=1}^N (\beta_1 + \beta_2 \rcuti - C(g_i))^2 \text{.}
\end{equation}
Now we need to minimize two parameters, $\beta_1$ and $\beta_2$, which yields a system of two equations that we write in matrix form:
\begin{align}
&R \vec{\beta}
= \vec{\epsilon}
\text{,}
\qquad 
R =
\begin{pmatrix}
N & \sum_i \rcuti \\
\sum_i \rcuti & \sum_i (\rcuti)^2
\end{pmatrix}
\text{,}
\\\nonumber&
\vec{\beta} =
\begin{pmatrix}
\beta_1 \\
\beta_2
\end{pmatrix}
\text{,}
\qquad
\vec{\epsilon} =
\begin{pmatrix}
\sum_i C(g_i) \\
\sum_i \rcuti C(g_i)
\end{pmatrix}
\text{.}
\end{align}
Using the linearity of the convolution, we define
$\vec{\gamma}$ through
\begin{align}
\vec{\epsilon} &=
\begin{pmatrix}
\sum_i C(g_i) \\
\sum_i \rcuti C(g_i)
\end{pmatrix}
=
\begin{pmatrix}
C(\sum_i g_i) \\
C(\sum_i \rcuti g_i)
\end{pmatrix} \\\nonumber
&=
C
\begin{pmatrix}
\gamma_1 \\
\gamma_2
\end{pmatrix}
=C(\vec{\gamma})\,\text{,}
\end{align}
where the convolution is understood to act on each component if applied on a vector.

This form, \mbox{$R \vec{\beta} = \vec{\epsilon}$}, holds true for any polynomial, with matrices $R$ and vectors $\vec{\beta}, \vec{\epsilon}$, whose dimensions are determined by the degree $d$ of the polynomial.
We determine $\vec{\beta}$ by inverting $R$, so that \cref{eq:constant-fit-result} generalizes to
\begin{equation}
\vec{\beta} = R^{-1} \vec{\epsilon} \text{.}
\end{equation}
We find the polynomial generalization of \cref{eq:grid-fit},
\begin{equation}
\vec{\tilde{g}} = R^{-1} \vec{\gamma} \text{,}
\end{equation}
where $\vec{\tilde{g}}, \vec{\gamma}$ are column vectors of grids and $R^{-1}$ a matrix with real numbers.
To extrapolate $\sigma$ in \mbox{$\rcut \to 0$}, we simply take the first component:
\begin{equation}
\sigma (\rcut = 0) = C ( \tilde{g}_1 ) = C \left( \sum_{k=1}^d R^{-1}_{1k} \gamma_k \right) \text{.}
\end{equation}

\subsection{Relic PDF dependence of grids from \MatrixHawaii{}}
Although an \emph{unweighted} least-squares fit would, in principle, allow for
a fully PDF-independent extrapolation of the interpolation grids,
a more reliable description of the \mbox{$\rcut\to0$} behaviour is typically
achieved when the statistical uncertainties $\Delta\sigma (\rcuti)$
of the cross sections $\sigma (\rcuti)$ are included, i.e.\ by
performing a \emph{weighted} least-squares fit.
Because \PineAPPL{} grids currently do not store statistical integration errors,
we use those provided by the direct \Matrix{} computation, which is based on a
specific PDF set. However, the differences in the uncertainties at various
$\rcuti$ mainly stem from numerical difficulties near the \mbox{$\rcut\to0$} limit.
Therefore, assuming that no significant PDF dependence is introduced at
this step --- which is supported by numerical checks --- is well justified.

Explicitly, in the simplified example of the fit of a constant,
\cref{eq:constant-fit-unweighted-leastsquares} becomes for a weighted
least-squares fit:
\begin{equation}
S = \sum_{i=1}^N \frac{(\beta - \sigma (\rcuti))^2}{(\Delta\sigma (\rcuti))^2}\, \text{.}
\label{eq:constant-fit-weighted-leastsquares}
\end{equation}
Instead of \cref{eq:constant-fit-result}, \mbox{$\partial S/\partial \beta = 0$} then yields:
\begin{equation}
\beta = \sum_{i=1}^N \underbrace{\frac{(\Delta\sigma (\rcuti))^{-2}}{\sum_{j=1}^N(\Delta\sigma (\rcutj))^{-2}}}_{=:w_i}\,\sigma (\rcuti)\, \text{,}
\label{eq:constant-fit-result-weighted}
\end{equation}
and the resulting interpolation grid reads
\begin{equation}
\tilde{g} = \sum_{i=1}^N w_ig_i\, \text{,}
\label{eq:grid-fit-weighted}
\end{equation}
which can be considered PDF-independent only if this assumption also holds for the weights 
$w_i$.

An equivalent argument applies to more general fits.
For the quadratic fit employed in \MatrixHawaii{}, we have verified numerically that any
PDF dependence of the extrapolation is, in practice, negligible.

In \MatrixHawaii{} the extrapolated grid, $\tilde g$, is provided as the default output to the user, but all the intermediate grids at fixed $\rcut$ values are still stored in the run folder.
As any \PineAPPL{} interpolation grids, they do not contain explicit error information. However, the combined integration plus extrapolation errors for the reference PDF set, as provided by \Matrix{}, is supplied in the metadata of the output grid.

\bibliography{matrix_alle_hawaii}

\end{document}